\documentclass{optica-article}

\journal{opticajournal} 

\articletype{Research Article}

\usepackage{pdfpages}

\usepackage{lineno}

\begin{document}

\title{Dose-efficient Quantum Phase Estimation in Lossy Optical Interferometry}

\author{Qilin Yu,\authormark{1} Ben Wang,\authormark{1,2,$\dag$} Kaimin Zheng,\authormark{1} Minghao Mi,\authormark{1} Hui Li,\authormark{1} and Lijian Zhang\authormark{1,*}}

\address{\authormark{1}National Laboratory of Solid State Microstructures and College of Engineering and Applied Sciences, Nanjing University, Nanjing 210093, China
}
\address{\authormark{2} School of Materials Science and Physics, China University of Mining and Technology, Xuzhou 221116, China}

\email{\authormark{$\dag$}ben.wang@cumt.edu.cn }
\email{\authormark{*}Corresponding author: lijian.zhang@nju.edu.cn}

\begin{abstract*}
Optical interferometry is a cornerstone technique for precise phase measurements across various fields. In many applications, for example biological imaging, it often necessitates stringent limits on light intensity to prevent adverse effects on light-sensitive samples, a condition known as dose-limited regimes.
Maximizing the precision per dose is therefore crucial.
In quantum metrology, quantum correlations enable high precision in phase estimation while adhering to dose constraints. 
Nevertheless, photon loss, including absorption by a sample, substantially diminishes the benefits of quantum enhancement in interferometry.
In this work, we experimentally investigate a dose-efficient approach to quantum phase estimation using sequential strategies in the presence of loss.
Performance of sequential strategies with and without control is evaluated through quantum Fisher information (QFI) per dose.
Experimental results show that both sequential strategies exceed the classical limit and outperform the parallel strategy using unbalanced N00N states. Notably, the control-enhanced sequential strategy attains superior QFI per dose, approaching the quantum limit. 
These results highlight the promise of sequential strategy for imaging and sensing in resource-constrained scenarios, marking a significant step toward practical and efficient quantum metrology in lossy environments.
\end{abstract*}

\section{Introduction}
Optical interferometry for measuring phase $\phi$ is an established tool across fields such as astronomy and biology. Quantum metrology enhances phase estimation precision by harnessing quantum resources~\cite{PhysRevLett.71.1355,doi:10.1126/science.1104149}, with noteworthy applications including gravitational wave detection~\cite{PhysRevA.88.041802} and clock synchronization~\cite{PhysRevA.72.042301}. 
The phase sensitivity in interferometry can surpass the shot-noise limit by leveraging quantum resources such as squeezing~\cite{PhysRevLett.130.073601,PhysRevA.81.033819,PhysRevLett.130.123603} and entanglement~\cite{Liu2021,PhysRevLett.118.233603}. 
The parallel strategy employing entangled N00N states~\cite{Walther2004,Mitchell2004} offers quantum-enhanced precision under ideal conditions.
However, the generation and preservation of high-N00N states are extremely demanding, hindering their usability in real-world applications~\cite{Mitchell2004,Dowling2008,doi:10.1126/science.1188172}. 
Alternatively, the standard multi-pass sequential strategy, where a single probe interacts coherently with the sample multiple times~\cite{PhysRevLett.96.010401,Higgins2007,PhysRevA.80.052114}, has demonstrated optimal performance in many scenarios~\cite{Juffmann2017,Juffmann2016,PhysRevLett.123.040501,doi:10.1126/sciadv.abd2986}.
Incorporating control or adaptive feedback within the sequential strategy offers further advantages, such as parameter estimation to Heisenberg limit under general dynamics~\cite{PhysRevLett.123.040501,PhysRevLett.115.110401,doi:10.1126/sciadv.abd2986} and optimal performance under realistic conditions~\cite{https://doi.org/10.1002/qute.202100080,PhysRevA.96.012117}.

Although both parallel and sequential strategies offer theoretical quantum advantages, their practical implementation faces significant challenges~\cite{10.1063/1.4724105,Ono2013,Casacio2021,Tran_ele_misco}. In photonic systems, imperfections such as decoherence and photon loss substantially degrade the enhanced precision~\cite{Escher2011NoisySystem,PhysRevA.79.023812,Kacprowicz2010}.
Additionally, in many applications like the sensing of light-sensitive samples, the estimation precision is constrained by the maximum light intensity the sample can tolerate without damage—a limit often determined by complex biochemical factors~\cite{Taylor2013,Morris2015,Moreau2017,doi:10.1126/science.aau1044,He2023}.
This trade-off between precision and sample safety underscores the need for sensing strategies that maximize the information extracted relative to the actual number of photons that interact with the sample.

We define a single experimental trial as a complete interrogation-and-readout cycle, which yields a per-trial quantum Fisher information (QFI) $F$ and consumes a per-trial dose $d$. Here, the dose $d$ represents the expected number of photon-sample interactions occurring per trial, accounting for all photons that enter the sample channel regardless of whether they are subsequently detected or lost.
As an example, for a probe state $\alpha |1 0\rangle+\sqrt{1-\alpha^2} |01\rangle$, where $|10\rangle$ denotes one photon in the sample mode and zero in the reference mode, the per-trial dose is $d=|\alpha|^2$.
For a sequence of $\mu$ independent trials, both the QFI and the dose are additive, yielding a total QFI of $F_{\text{tot}} = \mu F$ and a total dose of $D = \mu d$.
In the dose-limited regime, it is natural to define the QFI per dose \cite{koppell2024optimaldoselimitedphaseestimation}, 
\begin{equation}
    \xi:=\frac{F}{d},
\end{equation}
which quantifies the information extracted per unit photon–sample interaction.
Under a given total dose constraint $D \leq D_{\text{th}}$, the attainable precision is governed by the quantum Cramér–Rao bound~\cite{cramer1999mathematical},
$\delta \hat{\phi} \geq 1/\sqrt{D_{\mathrm{th}}\, \xi}$. Consequently, maximizing the achievable precision is equivalent to maximizing $\xi$.
Complementing standard metrics that evaluate precision per trial or per detected photon, the QFI per dose $\xi$ provides a physically motivated alternative metric for lossy optical interferometry by incorporating lost photons that have already interacted with the sample.
Sequential strategies, particularly those incorporating control, offer a promising route to enhance $\xi$ and approach the ultimate dose-limited bounds. 
However, their experimental realization under realistic lossy conditions remains largely elusive. Developing and implementing such dose-efficient quantum strategies is therefore both a key challenge and a major opportunity for advancing precision metrology in light-sensitive systems.

In this work, we experimentally investigate the performance of sequential strategies, including 
a passive multi-pass (MP) strategy and a control-enhanced sequential (CS) strategy.
Our results underscore the metrological advantages of sequential strategies, with surpassing the classical limit in phase estimation.
Moreover, despite the presence of loss, both sequential strategies achieve enhanced precision per dose compared to the theoretical benchmark of parallel N00N-state strategy.
Remarkably, CS strategy consistently attains higher QFI per dose than MP strategy and closely approaches the fundamental quantum limit, demonstrating a clear quantum advantage in a regime directly relevant to weak-beam sensing applications. These findings provide a practical and resource-efficient pathway toward optimal metrology in light-sensitive systems.

\begin{figure}[htbp]
    \centering
\includegraphics[width=0.8\textwidth]{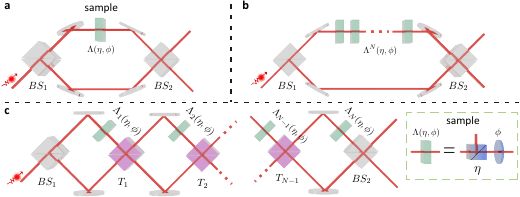} \caption{\label{fig:schematic fig}~a, The single pass (SP) strategy. b, The multi-pass (MP) strategy, where photons interact with the sample $N$ times. c, The $N$-pass control-enhanced sequential (CS) strategy, featuring adjustable beam-splitters (BSs) denoted as $T_k (k=1,...,N-1)$ as control modules. The sample, modeled by a BS of transmissivity $\eta$ and a phase shifter $\phi$, is illustrated in the green dashed box. In all strategies, the input single photon is partitioned between two modes by the BS$_1$ with transmissivity $\alpha^2$. After single or multiple interactions with the sample, the beams are redirected to the BS$_2$ with transmissivity 0.5, and photon detection occurs at the output ports.
}
\end{figure}

\section{Theoretical framework}

We consider a two-arm Mach-Zehnder interferometer (MZI) with a lossy sample placed in the upper arm, which is modeled using a beam-splitter (BS) and a phase shifter, as illustrated in Fig.\ref{fig:schematic fig}. 
Most of the loss occurs during the light-sample interaction that contributes to phase accumulation~\cite{PhysRevLett.102.040403,PhysRevA.80.013825}. 
The action of the sample is represented as $\Lambda(\eta,\phi)$, where $\eta$ denotes the transmissivity of the sample and $\phi$ represents the relative phase between the upper and lower arms.
The figure of merit that we aim to maximize is the QFI per dose $\xi=F/d$, where $d$ denotes the average dose interacting with the sample per experimental trial, and $F$ represents the per-trial QFI.
For pure states $\vert \zeta(\phi) \rangle$, the QFI is given by $F=4(\langle \partial_{\phi} \zeta (\phi)\vert \partial_{\phi} \zeta (\phi)\rangle - \vert \langle \partial_{\phi} \zeta (\phi)\vert \zeta(\phi)\rangle \vert^2 )$~\cite{Toth_2014}.
In phase estimation in the presence of loss, there exists an ultimate upper bound for single-arm loss with transmissivity $\eta$:
$F\leq  4\eta \bar{n}/(1-\eta)$,
where $\bar{n}$ is the mean photon number incident on the sample arm
~\cite{Escher2011,demkowicz2012elusive,PhysRevA.82.053804,OhnoBezerra_2025}.
Notably, this bound can be asymptotically saturated (as $\bar{n}\rightarrow \infty$) with appropriate interferometric schemes using displaced, weakly squeezed states ~\cite{PhysRevA.88.041802,Demkowicz2015}.
By normalizing the QFI bound against the corresponding dose, we define the quantum limit (QL) of the QFI per dose as
\begin{equation}
\xi_{\mathrm{QL}}\equiv  \frac{4 \eta}{1-\eta}.
\label{eq:QL}
\end{equation}
In practical sensing applications, if the sample can tolerate a maximum total dose $D_{\mathrm{th}}$, then the total QFI is strictly upper-bounded by $\xi_{\mathrm{QL}} D_{\mathrm{th}}$.
Consequently, the ultimate lower bound on estimation uncertainty is given by $\delta\phi_{\text{QL}} = 1/\sqrt{\xi_{\text{QL}}D_{\text{th}} }$.

We consider an initial two-mode state in the interferometer with a well-defined photon number $N$, given by $\vert \Psi \rangle =\sum_{k=0}^N \alpha_k \vert k,N-k \rangle$, where $\vert N_1, N_2 \rangle$ denotes a two-mode Fock state with $N_1$ and $N_2$ photons in the upper and lower arms, respectively.
After passing the sample and tracing out the ancillary mode, the output state is given by
$\rho(\phi) = \bigoplus_{l=0}^N p_l \vert \zeta_l (\phi) \rangle \langle  \zeta_l (\phi) \vert$, where 
$\vert \zeta_l(\phi) \rangle$ 
is the conditional pure state corresponding to the event when $l$ photons are lost in upper mode, and $p_l$ is the normalization factor corresponding to the probability of that event. 
We can then get the QFI and QFI per dose from $\rho(\phi)$ (See supplement 1 in~\cite{supp}).

A prominent two-mode entangled state is the balanced N00N state, $(\vert N0 \rangle + \vert 0N \rangle)/\sqrt{2}$, which achieves Heisenberg limit under ideal conditions but is highly fragile to loss~\cite{lossinducedlimitsPRA,Gilbert:08}. Nevertheless, it is worthwhile to investigate optimized, unbalanced N00N states, expressed as $\alpha \vert N0 \rangle + \sqrt{1-\alpha^2} \vert 0N \rangle$, as they potentially provide greater robustness against loss~\cite{PhysRevA.80.013825}. In lossy interferometry,
the QFI per dose for unbalanced N00N states is 
$ \xi_{\mathrm{unb}}=4 \eta^N {N} {(1-\alpha^2)}/{(1-\alpha^2+\eta^N \alpha^2)} $, which decreases as $\alpha^2$ increases. 
Treating $N$ as a continuous variable to determine the analytical upper bound, the optimized ratio for $\alpha \rightarrow 0$ and $\eta\rightarrow 1$ is $(\xi_{\mathrm{unb}}/\xi_{\mathrm{QL}})_{\mathrm{opt}} =(1 -1/\eta)/(e \ln \eta)$, occurring at the continuous optimum $N^*_{\mathrm{unb}} = -1/\ln\eta$.
In practice, where the photon number is an integer, the optimal photon number is $N_\mathrm{unb,opt}=\arg \max_{N \in \{ \lfloor N^*_\mathrm{unb} \rfloor, \lceil N^*_\mathrm{unb}\rceil\} } \xi_\mathrm{unb} (N)$, where $\lfloor \cdot \rfloor, \, \lceil \cdot \rceil$ denote the floor and ceiling functions, respectively (See supplement 2 in~\cite{supp}).

In the following analysis, we investigate sequential strategies (Fig.\ref{fig:schematic fig}~a-c) utilizing single photons. Single photons are injected into one port of BS$_1$ with transmissivity $\alpha^2$, which acts as $\vert 10\rangle \rightarrow -\sqrt{1-\alpha^2} \vert 10 \rangle +\alpha \vert 01 \rangle $ and $ \vert 01\rangle \rightarrow \alpha \vert 10 \rangle + \sqrt{1-\alpha^2} \vert 01 \rangle$, where, without loss of generality, $\alpha$ is taken to be real.
The single-pass (SP) strategy is shown in Fig.\ref{fig:schematic fig}~a. 
After passing through BS$_1$, a single photon will be in a superposition of two modes
$ \vert \psi \rangle = \alpha \vert 10\rangle + \sqrt{1- \alpha ^2} \vert 01\rangle $, which is a special case of unbalanced N00N states ($N$=1).
The QFI per dose can be optimized as $\xi_{\mathrm{SP}} =4\eta$, when $\alpha \rightarrow 0$ (See supplement 3 in~\cite{supp}). The classical limit (CL) is obtained by sending independent photons one-by-one through the interferometer, which is in accordance with that using coherent state~\cite{PhysRevA.80.063803}.
 
In the MP strategy (Fig.\ref{fig:schematic fig}~b), a single photon $\vert \psi \rangle$ traverses the sample $N$ times with the surviving probability of no loss across all passes. This sequential passage amplifies the accumulated phase by a factor of $N$, similar to unbalanced N00N states, thereby achieving the same QFI.
However, the dose imposed on the sample in the MP strategy is different due to the loss of the sample, $d_{\mathrm{MP}} = \alpha^2 \sum_{k=0}^{N-1} \eta^{k}$. Consequently, the MP strategy offers improved QFI per dose by achieving the same QFI with a lower dose. The QFI per dose is
\begin{equation}
\xi_{\mathrm{MP}} = 4 {\eta^N} N^2 \frac{1-\eta}{1-\eta^N} \left( 1- \frac{\eta^N \alpha^2}{ 1-\alpha^2 +\eta^N \alpha^2 }\right),
\label{eq:MP}
\end{equation} 
which decreases as $\alpha^2$ increases.
Under the continuous $N$ approximation, the analytical upper bound in the limit $\alpha \to 0$ is $(\xi_{\mathrm{MP}} / \xi_{\mathrm{QL}} )_{\mathrm{opt}}\approx 0.648 (1-\eta )/(\eta \ln ^2 \eta)$, attained at $ N^*_{\mathrm{MP}} \approx -1.6/\ln \eta$.
For discrete implementation, the optimal pass number $N_{\mathrm{MP,opt}} = \arg \max_{N \in\{ \lfloor N^*_\mathrm{MP}\rfloor, \lceil N^*_\mathrm{MP}\rceil \}} \xi_\mathrm{MP} (N)$ (See supplement 4 in~\cite{supp}).
For the highly transmissive samples ($\eta \rightarrow 1)$, the MP strategy can be optimized by choosing $N_{\mathrm{MP,opt}} \gg 1$.

In the sequential strategy, control can be helpful. One may expect an enhanced precision if the intermediate state can be adjusted before it re-enters the sample for the next pass. The refinement leads to the improved CS strategy, as shown in Fig.\ref{fig:schematic fig}~c. In the CS strategy, each of $N$ stages includes a control operation that couples the sample (up) and reference (down) modes via a BS with adjustable transmissivity. Let $T_k$ represent the $k$-th adjustable BS with a transmissivity coefficient $t_k$ ($1 \leq k \leq N-1$).
The initial state after the BS$_1$ is $\vert \psi^{(0)} \rangle = \alpha_0\vert 01 \rangle + \alpha_1\vert 10 \rangle$, where $\alpha_{0}=\sqrt{1-\alpha^2}$ and $\alpha_1= \alpha$ are the amplitudes of the reference and sample modes, respectively.
Due to photon loss, the final density matrix is a mixture of a survived single-photon component and a vacuum component.
Since the vacuum state carries no phase information, we focus on the post-selected pure state $\vert \psi^{(N)} \rangle$ conditioned on the survival of the photon:
\begin{equation}\vert \psi^{(N)} \rangle = \frac{\hat{\mathcal{M}}_N \vert \psi^{(0)} \rangle} {\sqrt{\langle \psi^{(0)} \vert \hat{\mathcal{M}}_N^\dagger \hat{\mathcal{M}}_N \vert \psi^{(0)} \rangle}} 
= \frac{\alpha_0^{(N)} \vert 01 \rangle + \alpha_1^{(N)} \vert 10 \rangle}{\sqrt{P^{(N)}}},
\end{equation}
where $\hat{\mathcal{M}}_N = \Lambda(\eta, \phi)\prod_{k=N-1}^1 [T_k \Lambda(\eta, \phi)]$ denotes the cumulative Kraus operator for the zero-loss trajectory. We also consider using the identical control in each round, $T_k=T$.
The survival probability is $P^{(N)} = \vert \alpha_0^{(N)} \vert^2 + \vert \alpha_1^{(N)} \vert^2$, where $\alpha_0^{(N)}$ and $\alpha_1^{(N)}$ are the unnormalized amplitudes after $N$ passes, with the initial amplitude $\alpha_1^{(0)}=\alpha$.
The per-trial dose, representing the expectation of the number of photon-sample interactions, is $d_{\mathrm{CS}} = \sum_{k=0}^{N-1} \vert \alpha^{(k)}_1 \vert^2$.
To evaluate the ultimate performance of the CS strategy, the fundamental figure of merit is the QFI per dose, $\xi_{\mathrm{CS}} = F_{\mathrm{CS}}/d_{\mathrm{CS}}$.
The per-trial QFI $F_{\mathrm{CS}}$ consists of two distinct parts: the classical information $F_{\text{cl}}$, originating from the sensitivity of the survival probability to $\phi$, and the post-selected information $F_{\text{post}}$, arising from the phase evolution of the surviving photon.
The QFI per dose $\xi_\mathrm{CS}$ attains its maximum at $\phi = 0$. At this optimal operating point, the first-order derivative $\partial_\phi P^{(N)}$ identically vanishes, rendering $F_{\text{cl}}=0$, and the per-trial QFI reduces strictly to the pure-state contribution,
$F_{\mathrm{CS}} =F_{\mathrm{post}} =
4 P^{(N)} [  \langle  \partial_{\phi} { \psi} ^{(N)} \vert  \partial_{\phi} { \psi} ^{(N)}  \rangle - 
\vert  \langle  \partial_{\phi} { \psi} ^{(N)} \vert { \psi} ^{(N)} \rangle  \vert ^2  ]$.
By utilizing this $N$-stage CS strategy for a sample with transmissivity $\eta$, we numerically optimize the QFI per dose $\xi_{\mathrm{CS}}$. The results demonstrate that the CS strategy achieves $\xi_{\mathrm{CS}}/\xi_{\mathrm{QL}}>0.9$ across the entire loss regime $\eta \in (0,1)$ (See supplement 5 in~\cite{supp}).

For single-photon superposition output states in MP and CS strategies, the optimal measurement to achieve the QFI for certain phases involves interfering with the two modes on a 50:50 beam-splitter (BS$_2$ in Fig.\ref{fig:schematic fig}).

\begin{figure}[t]
    \centering
\includegraphics[width=0.7\textwidth]{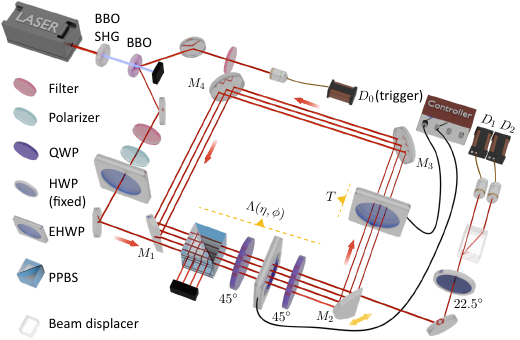}
\caption{\label{fig:expsetup}Experimental setup.
Photon pairs are generated through the parametric downconversion process in a $\beta$-barium borate (BBO) crystal pumped by frequency-doubled Ti:Sapphire laser pulses. The idler photon serves as a trigger ($D_0$). 
The heralded single photons undergo the evolution $\Lambda(\eta,\phi)$ and the control $T$ multiple times in the loop composed of four mirrors $M_{1-4}$, where $M_1$ and $M_2$ are D-shape mirrors. The number of passes depends on the position of $M_2$.
Finally, the photons are collected with single-mode fibers and detected using two single-photon detectors, $D_1$ and $D_2$.}
\end{figure}

\begin{figure*}[t]
    \centering
\includegraphics[width=\textwidth]
{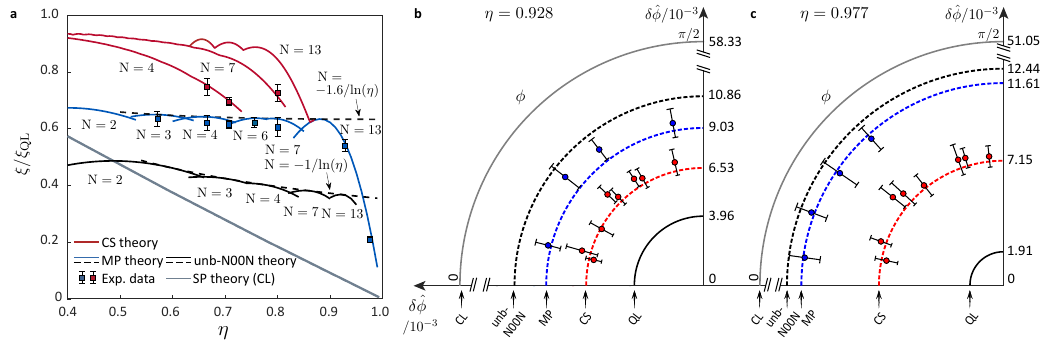}
\caption{a, QFI per dose, $\xi$, relative to the QL $\xi_{QL}$ for estimation of a single lossy phase with strategies: a single-pass (SP) strategy with a classical probe (gray dashed line), the parallel strategy with $N$-photon unbalanced N00N states (black dashed and solid lines), the MP strategy with $N$ passes, the CS strategy with $N$-stage optimal controls. Here the solid lines demonstrate the ratios of QFI per dose for the strategies when $N$ are integer numbers, and dashed lines demonstrate the optimal ratios for the strategies with continuous $N$.
b-c, Phase estimation uncertainty $\delta \hat{\phi} = 1/\sqrt{D_{\text{th}}\xi}$ of different $\phi \in (0,\pi/2)$ with $N = 13$ for $\eta = \{0.928, 0.977\}$ and $D_\text{th}=\{1231, 1602\}$, respectively. The radial direction represents $\delta \hat{\phi}$, and the angular direction (clockwise) represents the measured phase $\hat{\phi}$.}
\label{fig:eta_p_delta_phi}
\end{figure*} 

\section{Experimental setup and results}
The experimental setup is shown in~Fig.\ref{fig:expsetup}.
We employ a polarization MZI to estimate an unknown phase, using horizontal and vertical polarizations as the two modes.
Heralded single photons are initialized in the horizontal polarization state. An electronically controlled half-wave plate (EHWP) prepares the state as $\vert \psi \rangle = \alpha \vert 1_H 0_V\rangle + \sqrt{1-\alpha^2}\vert 0_H 1_V \rangle$.
A lossy sample with a variable phase is simulated using a partially polarizing beam-splitter (PPBS), two quarter-wave plates (QWPs), and an EHWP. The PPBS has transmissivities $\eta_H$ and $\eta_V$ for horizontal and vertical polarizations, respectively, with the relative transmissivity defined as $\eta=\eta_V/\eta_H$ ($\eta_V<\eta_H$).
Experimentally, $\eta$ is calibrated to account for the total loss in the interferometer~(See supplement 6 in~\cite{supp}).
An adjustable EHWP and two $45^{\circ}$ QWPs introduce a phase $\phi$ between two modes.
Another EHWP implements the control $T$ to redistribute the intensity between two polarization modes. 
Multiple passes are achieved via a four-mirror loop, $M_{1-4}$~(See supplement 6 in~\cite{supp}). 
The optimal measurement is realized with the $22.5^{\circ}$ HWP and a birefringent calcite beam displacer, followed by two single-photon detectors.

To illustrate the performance of sequential strategies described above, we investigate QFI per dose versus the QL, $\xi/\xi_{\mathrm{QL}}$.
We implement $N$-pass MP and CS strategies to estimate different phases $\phi \in (0,\pi/2)$ with different $\eta\in (0.5,1)$.
For each $\eta$, we perform a scan over the $\pi/N$ period to get the likelihood function and then get the estimated value $\hat{\phi}$ using maximum likelihood estimation. With the phase estimation uncertainty $\delta ^2 \hat{\phi}$, we get the QFI per dose $\xi =1/ ({d}\, \delta ^2 \hat{\phi})$ (See supplement 7 in~\cite{supp}).

\subsection{(A) QFI per dose related to $N$}
We evaluate the QFI per dose with respect to different pass numbers $N$ under a fixed dose limit $D_{\text{th}}=195$ with $\alpha^2=0.1$.
In Fig.\ref{fig:eta_p_delta_phi}~a, the black dashed lines represent the analytical upper bounds, yielding $(\xi_{\mathrm{unb}}/\xi_{\mathrm{QL}})_{\mathrm{opt}} \approx 0.38$ for the unbalanced N00N benchmark and $(\xi_{\mathrm{MP}}/\xi_{\mathrm{QL}})_{\mathrm{opt}} \approx 0.65$ for the MP strategy, evaluated at their respective continuous optima $N^*_{\text{unb}}$ and $N^*_{\text{MP}}$ when $\eta\approx 1$. The solid colored lines illustrate the theoretical predictions for $\xi/\xi_{\mathrm{QL}}$ at fixed discrete values of $N$ as a function of the continuous transmissivity $\eta$. In particular, $N \in \{2, 3, 4, 6, 7, 13\}$ for the MP strategy (blue curves), and $N \in \{4, 7, 13\}$ for the CS strategy (red curves). For comparison, we also plot the theoretical results for the SP strategy (gray line) and the parallel strategy using unbalanced N00N states (black curves). Experimentally, we implement $N \in \{3, 4, 6, 7, 13\}$ for the MP strategy and $N \in \{4, 7\}$ for the CS strategy across representative $\eta \in \{0.565, 0.665, 0.714, 0.752, 0.794, 0.928, 0.977\}$. For the MP strategy, the measured $\xi/\xi_{\mathrm{QL}}$ values show close agreement with the discrete theoretical predictions within error bars when the chosen $N=\{3,4,6,7\} \approx N_{\text{MP,opt}}$.
Under the same $(\eta,N)$, CS strategy consistently yields larger QFI per dose than MP strategy due to the optimized control, i.e., $\xi_{\mathrm{CS}} \geq \xi_{\mathrm{MP}} $. 
Despite $N$ being smaller than the optimal value for a given $\eta$, the improvement is already evident.

\subsection{(B) Phase estimation uncertainty}
We fix the total dose limit $D_{\text{th}}$ and implement the MP and CS strategies with the pass number $N=13$ to estimate phases $\phi \in (0,\pi/2)$.
Fig.\ref{fig:eta_p_delta_phi}~b-c shows the resulting phase uncertainty $\delta \hat{\phi} = 1/\sqrt{D_{\text{th}} \xi}$ as a polar plot.
Under the same $(\eta,\,D_\text{th})$, we compare MP strategy (blue dashed circular arcs and data) and CS strategy (red dashed circular arcs and data) with the parallel benchmark given by 13-photon unbalanced N00N states (black dashed circular arcs). The QL is indicated by black solid circular arcs, and the CL by gray solid circular arcs. 
For $\eta=0.928$ and $D_{\text{th}}$\;=\;1231 (Fig.\ref{fig:eta_p_delta_phi}~b), we use 150 photons in MP strategy to estimate phases $\hat{\phi}$\;=\;$\{$0.250, 0.665, 0.912, 1.386$\}$ rad, and 180 photons in CS strategy to estimate phases $\hat{\phi}$\;=\;$\{$0.223, 0.279, 0.758, 0.804, 0.992, 1.056, 1.347$\}$ rad. For $\eta=0.977$ and $D_{\text{th}}=1602$ (Fig.\ref{fig:eta_p_delta_phi}~c), we use 150 photons in MP strategy to estimate phases $\hat{\phi}$\;=\;$\{$0.131, 0.363, 0.608, 0.829$\}$\;rad, and 195 photons in CS strategy to estimate phases $\hat{\phi}$\;=\;$\{$0.206, 0.342, 0.673, 0.740, 0.899, 1.225, 1.281, 1.462$\}$\;rad. At fixed $(\eta,D_{\text{th}})$, MP strategy achieves smaller uncertainties than the parallel unbalanced N00N-state benchmark, consistent with the theoretical analysis in Eq.~(\ref{eq:MP}). 
While these estimated phases deviate from the optimal $\phi=0$, our measurement omits the classical contribution $F_{\text{cl}}$, yielding an underestimation of $\xi_\mathrm{CS}$. Notably, even with this conservative evaluation, the CS strategy leverages inter-pass control to further reduce $\delta \hat{\phi}$, surpassing the MP strategy and approaching the QL over the measured phase range.

\begin{figure}[hbtp]
\centering
\includegraphics{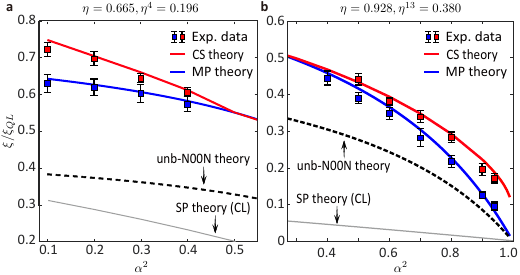}
\caption{Results for $\xi/\xi_{\mathrm{QL}}$ by varying $\alpha^2$: a, 4-stage MP and CS strategies at $\eta = 0.665$; b, 13-stage MP and CS strategies at $\eta= 0.928$. The results are compared to the parallel strategy using unbalanced N00N states and the SP strategy. \label{fig:p4p13_alpha}
}
\end{figure}

\subsection{(C) QFI per dose related to $\alpha^2$}
We study how the QFI per dose depends on the initial splitting $\alpha^2$ for two representative settings: a near-optimal pass number $N=4$ at $\eta=0.665$ under the total dose constraint $D_{\text{th}}=240$, and a larger pass number $N=13$ at $\eta=0.928$ under $D_{\text{th}}=1852.5$, as shown in Fig.\ref{fig:p4p13_alpha}~a-b.
Red curves and data correspond to the CS strategy, and blue curves and data correspond to the MP strategy. 
For reference, we plot the parallel unbalanced N00N-state benchmark and the SP strategy prediction.
In Fig.\ref{fig:p4p13_alpha}~a, the MP strategy reaches $\xi_{\text{MP}}/\xi_{\text{QL}}\approx 0.65$ at $\alpha^2=0.1$. As $\alpha^2$ increases, $\xi$ decreases. 
This is because larger $\alpha^2$ puts more weight in the lossy arm, so each traversal consumes more dose but yields less surviving, phase-sensitive amplitude. Consequently, the information gained per dose is reduced and $\xi$ drops monotonically with $\alpha^2$.
The CS strategy consistently outperforms the MP strategy and performs best in the weak-coupling regime $\alpha^2 \rightarrow 0$. 
In Fig.\ref{fig:p4p13_alpha}~b, neither MP nor CS strategy operates at its globally optimal $N$ at the given $\eta$. Even in this non-optimal $N$ setting, the CS strategy remains superior and produces a larger $\xi$ than the MP strategy across the same $\alpha^2$ range. 

\section{Discussion and conclusion}
Quantifying the enhancement achieved through quantum behavior requires a clear definition of resources.
In optical interferometry, the key resources are the number of input photons of the interferometer and the number of experimental trials required to achieve a given phase precision.
Under fixed total resources, parallel strategies employing $N$-photon N00N states and $N$-pass sequential strategies using single photons yield identical QFI.
However, in many practical scenarios such as biological imaging, photon availability is not the constraint-- sample damage or saturation thresholds (e.g., photobleaching or radiation damage) limit the photon dose that can interact with the sample. 
These interacting photons typically constitute only a small fraction of the total photons injecting into the interferometer, and it is this interacting fraction that sets the effective sensing resource~\cite{PhysRevLett.96.010401,Wiseman2009,PhysRevA.96.062109}.
Under a fixed total dose constraint $D_{\text{th}}$, the total obtainable information scales as $F_{\text{tot}}=D_{\text{th}} \xi$, so improving precision requires maximizing the QFI per dose $\xi$.
Sequential strategies incorporating the optimal control enable near-quantum-limited performance without exceeding the dose threshold.

While the QFI is the established gold standard for benchmarking precision limits, its operational significance requires a complete inference framework that ensures consistency with the estimator under finite resources. Our analysis relies on a local estimation regime where the likelihood is restricted to a single unambiguous interval, utilizing prior information to resolve the $2\pi/N$ phase ambiguity and bypassing the inferential overhead associated with global search~\cite{hradil2026}.
This ensures a fair comparison between the sequential strategies and parallel unbalanced N00N-state benchmark, as both protocols necessitate equivalent prior information to resolve phase ambiguities~\cite{Xiang2011}.
Although our demonstration focuses on local phase estimation, the framework is intrinsically compatible with both non-adaptive
~\cite{Higgins_2009,Full_period} and adaptive feedback schemes~\cite{Higgins2007,PhysRevA.57.2169,PhysRevA.65.043803,Zheng_kk20,Bonato2016,SwarmOptimization}, facilitating an extension to global phase estimation across the full $[0,2\pi)$ range.
While these global approaches are well-established for high-resource scenarios, the feasibility of maintaining global estimator consistency and the optimal dose allocation in the presence of loss~\cite{PhysRevA.82.053804} and dose constraints remain largely unexplored research frontiers.

In conclusion, we have experimentally investigated optical phase estimation under realistic loss using MP and CS strategies. Sequential strategies not only outperform the parallel unbalanced N00N-state benchmark, but also exhibit superior robustness against sample loss. Notably, the CS strategy achieves QFI per dose closely approaching the fundamental quantum limit.
Although demonstrated optically, the principles underlying the CS strategy are broadly applicable to quantum platforms where resources are constrained by probe exposure, energy, or particle number~\cite{Jin2013,squeezed_sequential2025}, rather than measurement time. 
Potential applications include microscopy~\cite{Koppell2019,https://doi.org/10.1002/lpor.201900097,PhysRevLett.130.056101}, spectroscopy~\cite{Kalashnikov2016} and nitrogen-vacancy (NV) center magnetometry ~\cite{Nusran_nanotech2012,QaunSen_Degen2017,Herbschleb2021} of radiation-sensitive samples.
Crucially, our framework is compatible with high-resolution phase imaging~\cite{Ono2013,PhysRevLett.112.103604,sciadv2021}. By employing a cavity-like configuration~\cite{Juffmann2016,Juffmann2017} or a recirculating loop with high-speed optical switches~\cite{PhysRevLett.129.150501}, all $N$ interactions can be confined to a collinear path at the same spatial coordinate. Full-field imaging can then be realized by scanning the sample point-by-point, with each pixel reconstructed using the dose-optimized sequential strategies.
By experimentally validating the CS strategy, our work establishes dose-efficient phase estimation as a general paradigm in quantum metrology and provides a framework for developing resource-efficient quantum sensors across diverse physical platforms.
 
\section{Back matter} 
\begin{backmatter}
\bmsection{Fundings}
Please see Acknowledgment for funding details.

\bmsection{Acknowledgment}
This work was supported by National Natural Science Foundation   of   China (Grants No. U24A2017, No. 12347104 and No. 12461160276), the National Key Research and Development Program of China (Grants No. 2023YFC2205802), Natural Science Foundation of Jiangsu Province (Grants No. BK20243060 and No. BK20233001).

\bmsection{Disclosures}
The authors declare that there are no conflicts of interest related to this article.
 
\bmsection{Data availability} Data underlying the results presented in this paper are available in Supplement (Ref.~\cite{supp}).

\bmsection{Supplemental document}
See Supplement (Ref. \cite{supp}) for supporting content.

\end{backmatter}


\begin{thebibliography}{10}
\newcommand{\enquote}[1]{``#1''}

\bibitem{PhysRevLett.71.1355}
M.~J. Holland and K.~Burnett, \enquote{Interferometric detection of optical
  phase shifts at the heisenberg limit,} {\protect\JournalTitle{Phys. Rev.
  Lett.}} \textbf{71}, 1355--1358 (1993).

\bibitem{doi:10.1126/science.1104149}
V.~Giovannetti, S.~Lloyd, and L.~Maccone, \enquote{Quantum-enhanced
  measurements: Beating the standard quantum limit,}
  {\protect\JournalTitle{Science}} \textbf{306}, 1330--1336 (2004).

\bibitem{PhysRevA.88.041802}
R.~Demkowicz-Dobrza\ifmmode~\acute{n}\else \'{n}\fi{}ski, K.~Banaszek, and
  R.~Schnabel, \enquote{Fundamental quantum interferometry bound for the
  squeezed-light-enhanced gravitational wave detector geo 600,}
  {\protect\JournalTitle{Phys. Rev. A}} \textbf{88}, 041802 (2013).

\bibitem{PhysRevA.72.042301}
M.~de~Burgh and S.~D. Bartlett, \enquote{Quantum methods for clock
  synchronization: Beating the standard quantum limit without entanglement,}
  {\protect\JournalTitle{Phys. Rev. A}} \textbf{72}, 042301 (2005).

\bibitem{PhysRevLett.130.073601}
W.~Huang, X.~Liang, B.~Zhu, \emph{et~al.}, \enquote{Protection of noise
  squeezing in a quantum interferometer with optimal resource allocation,}
  {\protect\JournalTitle{Phys. Rev. Lett.}} \textbf{130}, 073601 (2023).

\bibitem{PhysRevA.81.033819}
T.~Ono and H.~F. Hofmann, \enquote{Effects of photon losses on phase estimation
  near the heisenberg limit using coherent light and squeezed vacuum,}
  {\protect\JournalTitle{Phys. Rev. A}} \textbf{81}, 033819 (2010).

\bibitem{PhysRevLett.130.123603}
J.~A.~H. Nielsen, J.~S. Neergaard-Nielsen, T.~Gehring, and U.~L. Andersen,
  \enquote{Deterministic quantum phase estimation beyond n00n states,}
  {\protect\JournalTitle{Phys. Rev. Lett.}} \textbf{130}, 123603 (2023).

\bibitem{Liu2021}
L.-Z. Liu, Y.-Z. Zhang, Z.-D. Li, \emph{et~al.}, \enquote{Distributed quantum
  phase estimation with entangled photons,} {\protect\JournalTitle{Nature
  Photonics}} \textbf{15}, 137--142 (2021).

\bibitem{PhysRevLett.118.233603}
G.~Colangelo, F.~Martin~Ciurana, G.~Puentes, \emph{et~al.},
  \enquote{Entanglement-enhanced phase estimation without prior phase
  information,} {\protect\JournalTitle{Phys. Rev. Lett.}} \textbf{118}, 233603
  (2017).

\bibitem{Walther2004}
P.~Walther, J.-W. Pan, M.~Aspelmeyer, \emph{et~al.}, \enquote{De broglie
  wavelength of a non-local four-photon state,} {\protect\JournalTitle{Nature}}
  \textbf{429}, 158--161 (2004).

\bibitem{Mitchell2004}
M.~W. Mitchell, J.~S. Lundeen, and A.~M. Steinberg, \enquote{Super-resolving
  phase measurements with a multiphoton entangled state,}
  {\protect\JournalTitle{Nature}} \textbf{429}, 161--164 (2004).

\bibitem{Dowling2008}
J.~P. Dowling, \enquote{Quantum optical metrology -- the lowdown on high-n00n
  states,} {\protect\JournalTitle{Contemporary Physics}} \textbf{49}, 125--143
  (2008).

\bibitem{doi:10.1126/science.1188172}
I.~Afek, O.~Ambar, and Y.~Silberberg, \enquote{High-noon states by mixing
  quantum and classical light,} {\protect\JournalTitle{Science}} \textbf{328},
  879--881 (2010).

\bibitem{PhysRevLett.96.010401}
V.~Giovannetti, S.~Lloyd, and L.~Maccone, \enquote{Quantum metrology,}
  {\protect\JournalTitle{Phys. Rev. Lett.}} \textbf{96}, 010401 (2006).

\bibitem{Higgins2007}
B.~L. Higgins, D.~W. Berry, S.~D. Bartlett, \emph{et~al.},
  \enquote{Entanglement-free heisenberg-limited phase estimation,}
  {\protect\JournalTitle{Nature}} \textbf{450}, 393--396 (2007).

\bibitem{PhysRevA.80.052114}
D.~W. Berry, B.~L. Higgins, S.~D. Bartlett, \emph{et~al.}, \enquote{How to
  perform the most accurate possible phase measurements,}
  {\protect\JournalTitle{Phys. Rev. A}} \textbf{80}, 052114 (2009).

\bibitem{Juffmann2017}
T.~Juffmann, S.~A. Koppell, B.~B. Klopfer, \emph{et~al.}, \enquote{Multi-pass
  transmission electron microscopy,} {\protect\JournalTitle{Scientific
  Reports}} \textbf{7}, 1699 (2017).

\bibitem{Juffmann2016}
T.~Juffmann, B.~B. Klopfer, T.~L. Frankort, \emph{et~al.}, \enquote{Multi-pass
  microscopy,} {\protect\JournalTitle{Nature Communications}} \textbf{7}, 12858
  (2016).

\bibitem{PhysRevLett.123.040501}
Z.~Hou, R.-J. Wang, J.-F. Tang, \emph{et~al.}, \enquote{Control-enhanced
  sequential scheme for general quantum parameter estimation at the heisenberg
  limit,} {\protect\JournalTitle{Phys. Rev. Lett.}} \textbf{123}, 040501
  (2019).

\bibitem{doi:10.1126/sciadv.abd2986}
Z.~Hou, J.-F. Tang, H.~Chen, \emph{et~al.}, \enquote{Zero–trade-off
  multiparameter quantum estimation via simultaneously saturating multiple
  heisenberg uncertainty relations,} {\protect\JournalTitle{Science Advances}}
  \textbf{7}, eabd2986 (2021).

\bibitem{PhysRevLett.115.110401}
H.~Yuan and C.-H.~F. Fung, \enquote{Optimal feedback scheme and universal time
  scaling for hamiltonian parameter estimation,} {\protect\JournalTitle{Phys.
  Rev. Lett.}} \textbf{115}, 110401 (2015).

\bibitem{https://doi.org/10.1002/qute.202100080}
J.~Liu, M.~Zhang, H.~Chen, \emph{et~al.}, \enquote{Optimal scheme for quantum
  metrology,} {\protect\JournalTitle{Advanced Quantum Technologies}}
  \textbf{5}, 2100080 (2022).

\bibitem{PhysRevA.96.012117}
J.~Liu and H.~Yuan, \enquote{Quantum parameter estimation with optimal
  control,} {\protect\JournalTitle{Phys. Rev. A}} \textbf{96}, 012117 (2017).

\bibitem{10.1063/1.4724105}
A.~Crespi, M.~Lobino, J.~C.~F. Matthews, \emph{et~al.}, \enquote{Measuring
  protein concentration with entangled photons,} {\protect\JournalTitle{Applied
  Physics Letters}} \textbf{100}, 233704 (2012).

\bibitem{Ono2013}
T.~Ono, R.~Okamoto, and S.~Takeuchi, \enquote{An entanglement-enhanced
  microscope,} {\protect\JournalTitle{Nature Communications}} \textbf{4}, 2426
  (2013).

\bibitem{Casacio2021}
C.~A. Casacio, L.~S. Madsen, A.~Terrasson, \emph{et~al.},
  \enquote{Quantum-enhanced nonlinear microscopy,}
  {\protect\JournalTitle{Nature}} \textbf{594}, 201--206 (2021).

\bibitem{Tran_ele_misco}
S.~A. Koppell, Y.~Israel, A.~J. Bowman, \emph{et~al.}, \enquote{{Transmission
  electron microscopy at the quantum limit},} {\protect\JournalTitle{Applied
  Physics Letters}} \textbf{120}, 190502 (2022).

\bibitem{Escher2011NoisySystem}
B.~M. Escher, R.~L. de~Matos~Filho, and L.~Davidovich, \enquote{Quantum
  metrology for noisy systems,} {\protect\JournalTitle{Brazilian Journal of
  Physics}} \textbf{41}, 229--247 (2011).

\bibitem{PhysRevA.79.023812}
L.~Maccone and G.~De~Cillis, \enquote{Robust strategies for lossy quantum
  interferometry,} {\protect\JournalTitle{Phys. Rev. A}} \textbf{79}, 023812
  (2009).

\bibitem{Kacprowicz2010}
M.~Kacprowicz, R.~Demkowicz-Dobrza{\'{n}}ski, W.~Wasilewski, \emph{et~al.},
  \enquote{Experimental quantum-enhanced estimation of a lossy phase shift,}
  {\protect\JournalTitle{Nature Photonics}} \textbf{4}, 357--360 (2010).

\bibitem{Taylor2013}
M.~A. Taylor, J.~Janousek, V.~Daria, \emph{et~al.}, \enquote{Biological
  measurement beyond the quantum limit,} {\protect\JournalTitle{Nature
  Photonics}} \textbf{7}, 229--233 (2013).

\bibitem{Morris2015}
P.~A. Morris, R.~S. Aspden, J.~E.~C. Bell, \emph{et~al.}, \enquote{Imaging with
  a small number of photons,} {\protect\JournalTitle{Nature Communications}}
  \textbf{6}, 5913 (2015).

\bibitem{Moreau2017}
P.-A. Moreau, J.~Sabines-Chesterking, R.~Whittaker, \emph{et~al.},
  \enquote{Demonstrating an absolute quantum advantage in direct absorption
  measurement,} {\protect\JournalTitle{Scientific Reports}} \textbf{7}, 6256
  (2017).

\bibitem{doi:10.1126/science.aau1044}
Y.~M. Sigal, R.~Zhou, and X.~Zhuang, \enquote{Visualizing and discovering
  cellular structures with super-resolution microscopy,}
  {\protect\JournalTitle{Science}} \textbf{361}, 880--887 (2018).

\bibitem{He2023}
Z.~He, Y.~Zhang, X.~Tong, \emph{et~al.}, \enquote{Quantum microscopy of cells
  at the heisenberg limit,} {\protect\JournalTitle{Nature Communications}}
  \textbf{14}, 2441 (2023).

\bibitem{koppell2024optimaldoselimitedphaseestimation}
S.~A. Koppell and M.~A. Kasevich, \enquote{Optimal dose-limited phase
  estimation without entanglement,} arXiv:2203.10137v2 (2024).

\bibitem{cramer1999mathematical}
H.~Cram{\'e}r, \emph{Mathematical methods of statistics}, vol.~26 (Princeton
  university press, 1999).

\bibitem{PhysRevLett.102.040403}
U.~Dorner, R.~Demkowicz-Dobrzanski, B.~J. Smith, \emph{et~al.},
  \enquote{Optimal quantum phase estimation,} {\protect\JournalTitle{Phys. Rev.
  Lett.}} \textbf{102}, 040403 (2009).

\bibitem{PhysRevA.80.013825}
R.~Demkowicz-Dobrzanski, U.~Dorner, B.~J. Smith, \emph{et~al.},
  \enquote{Quantum phase estimation with lossy interferometers,}
  {\protect\JournalTitle{Phys. Rev. A}} \textbf{80}, 013825 (2009).

\bibitem{Toth_2014}
G.~Tóth and I.~Apellaniz, \enquote{Quantum metrology from a quantum
  information science perspective,} {\protect\JournalTitle{Journal of Physics
  A: Mathematical and Theoretical}} \textbf{47}, 424006 (2014).

\bibitem{Escher2011}
B.~M. Escher, R.~L. de~Matos~Filho, and L.~Davidovich, \enquote{General
  framework for estimating the ultimate precision limit in noisy
  quantum-enhanced metrology,} {\protect\JournalTitle{Nature Physics}}
  \textbf{7}, 406--411 (2011).

\bibitem{demkowicz2012elusive}
R.~Demkowicz-Dobrza{\'n}ski, J.~Ko{\l}ody{\'n}ski, and M.~Gu{\c{t}}{\u{a}},
  \enquote{The elusive heisenberg limit in quantum-enhanced metrology,}
  {\protect\JournalTitle{Nature communications}} \textbf{3}, 1063 (2012).

\bibitem{PhysRevA.82.053804}
J.~Ko\l{}ody\ifmmode~\acute{n}\else \'{n}\fi{}ski and
  R.~Demkowicz-Dobrza\ifmmode~\acute{n}\else \'{n}\fi{}ski, \enquote{Phase
  estimation without a priori phase knowledge in the presence of loss,}
  {\protect\JournalTitle{Phys. Rev. A}} \textbf{82}, 053804 (2010).

\bibitem{OhnoBezerra_2025}
M.~E. Ohno~Bezerra, F.~Albarelli, and R.~Demkowicz-Dobrzanski,
  \enquote{Simultaneous optical phase and loss estimation revisited:
  measurement and probe incompatibility,} {\protect\JournalTitle{Journal of
  Physics A: Mathematical and Theoretical}} \textbf{58}, 265303 (2025).

\bibitem{Demkowicz2015}
R.~Demkowicz-Dobrza{\'{n}}ski, M.~Jarzyna, and J.~Ko{\l}ody{\'{n}}ski,
  \emph{Chapter Four - Quantum Limits in Optical Interferometry} (Elsevier,
  2015), vol.~60, pp. 345--435.

\bibitem{supp}
\enquote{Supplementary materials}.

\bibitem{lossinducedlimitsPRA}
M.~A. Rubin and S.~Kaushik, \enquote{Loss-induced limits to phase measurement
  precision with maximally entangled states,} {\protect\JournalTitle{Phys. Rev.
  A}} \textbf{75}, 053805 (2007).

\bibitem{Gilbert:08}
G.~Gilbert, M.~Hamrick, and Y.~S. Weinstein, \enquote{Use of maximally
  entangled n-photon states for practical quantum interferometry,}
  {\protect\JournalTitle{J. Opt. Soc. Am. B}} \textbf{25}, 1336--1340 (2008).

\bibitem{PhysRevA.80.063803}
T.-W. Lee, S.~D. Huver, H.~Lee, \emph{et~al.}, \enquote{Optimization of quantum
  interferometric metrological sensors in the presence of photon loss,}
  {\protect\JournalTitle{Phys. Rev. A}} \textbf{80}, 063803 (2009).

\bibitem{Wiseman2009}
H.~M. Wiseman, D.~W. Berry, S.~D. Bartlett, \emph{et~al.}, \enquote{Adaptive
  measurements in the optical quantum information laboratory,}
  {\protect\JournalTitle{IEEE Journal of Selected Topics in Quantum
  Electronics}} \textbf{15}, 1661--1672 (2009).

\bibitem{PhysRevA.96.062109}
P.~M. Birchall, J.~L. O'Brien, J.~C.~F. Matthews, and H.~Cable,
  \enquote{Quantum-classical boundary for precision optical phase estimation,}
  {\protect\JournalTitle{Phys. Rev. A}} \textbf{96}, 062109 (2017).

\bibitem{hradil2026}
Z.~Hradil and J.~Řeháček, \enquote{A realistic framework for quantum sensing
  under finite resources,} arXiv:2603.08306 (2026).

\bibitem{Xiang2011}
G.~Y. Xiang, B.~L. Higgins, D.~W. Berry, \emph{et~al.},
  \enquote{Entanglement-enhanced measurement of a completely unknown optical
  phase,} {\protect\JournalTitle{Nature Photonics}} \textbf{5}, 43--47 (2011).

\bibitem{Higgins_2009}
B.~L. Higgins, D.~W. Berry, S.~D. Bartlett, \emph{et~al.},
  \enquote{Demonstrating heisenberg-limited unambiguous phase estimation
  without adaptive measurements,} {\protect\JournalTitle{New Journal of
  Physics}} \textbf{11}, 073023 (2009).

\bibitem{Full_period}
L.-Z. Liu, Y.-Y. Fei, Y.~Mao, \emph{et~al.}, \enquote{Full-period quantum phase
  estimation,} {\protect\JournalTitle{Phys. Rev. Lett.}} \textbf{130}, 120802
  (2023).

\bibitem{PhysRevA.57.2169}
H.~M. Wiseman and R.~B. Killip, \enquote{Adaptive single-shot phase
  measurements: The full quantum theory,} {\protect\JournalTitle{Phys. Rev. A}}
  \textbf{57}, 2169--2185 (1998).

\bibitem{PhysRevA.65.043803}
D.~W. Berry and H.~M. Wiseman, \enquote{Adaptive quantum measurements of a
  continuously varying phase,} {\protect\JournalTitle{Phys. Rev. A}}
  \textbf{65}, 043803 (2002).

\bibitem{Zheng_kk20}
K.~Zheng, M.~Mi, B.~Wang, \emph{et~al.}, \enquote{Quantum-enhanced stochastic
  phase estimation with the su(1,1) interferometer,}
  {\protect\JournalTitle{Photon. Res.}} \textbf{8}, 1653--1661 (2020).

\bibitem{Bonato2016}
C.~Bonato, M.~S. Blok, H.~T. Dinani, \emph{et~al.}, \enquote{Optimized quantum
  sensing with a single electron spin using real-time adaptive measurements,}
  {\protect\JournalTitle{Nature Nanotechnology}} \textbf{11}, 247--252 (2016).

\bibitem{SwarmOptimization}
A.~J.~F. Hayes and D.~W. Berry, \enquote{Swarm optimization for adaptive phase
  measurements with low visibility,} {\protect\JournalTitle{Phys. Rev. A}}
  \textbf{89}, 013838 (2014).
 
\bibitem{Jin2013}
X.-M. Jin, C.-Z. Peng, Y.~Deng, \emph{et~al.}, \enquote{Sequential path
  entanglement for quantum metrology,} {\protect\JournalTitle{Scientific
  Reports}} \textbf{3}, 1779 (2013).

\bibitem{squeezed_sequential2025}
Y.~Feng, Z.~Zeng, J.~Cheng, \emph{et~al.}, \enquote{Quantum-enhanced
  interferometer for multiphase sensing,} {\protect\JournalTitle{Phys. Rev.
  Lett.}} \textbf{135}, 183602 (2025).

\bibitem{Koppell2019}
S.~A. Koppell, M.~Mankos, A.~J. Bowman, \emph{et~al.}, \enquote{Design for a
  10 kev multi-pass transmission electron microscope,}
  {\protect\JournalTitle{Ultramicroscopy}} \textbf{207}, 112834 (2019).

\bibitem{https://doi.org/10.1002/lpor.201900097}
M.~Gilaberte~Basset, F.~Setzpfandt, F.~Steinlechner, \emph{et~al.},
  \enquote{Perspectives for applications of quantum imaging,}
  {\protect\JournalTitle{Laser \& Photonics Reviews}} \textbf{13}, 1900097
  (2019).

\bibitem{PhysRevLett.130.056101}
C.~Dwyer, \enquote{Quantum limits of transmission electron microscopy,}
  {\protect\JournalTitle{Phys. Rev. Lett.}} \textbf{130}, 056101 (2023).

\bibitem{Kalashnikov2016}
D.~A. Kalashnikov, A.~V. Paterova, S.~P. Kulik, and L.~A. Krivitsky,
  \enquote{Infrared spectroscopy with visible light,}
  {\protect\JournalTitle{Nature Photonics}} \textbf{10}, 98--101 (2016).

\bibitem{Nusran_nanotech2012}
N.~M. Nusran, M.~U. Momeen, and M.~V.~G. Dutt, \enquote{High-dynamic-range
  magnetometry with a single electronic spin in diamond,}
  {\protect\JournalTitle{Nature Nanotechnology}} \textbf{7}, 109--113 (2012).

\bibitem{QaunSen_Degen2017}
C.~L. Degen, F.~Reinhard, and P.~Cappellaro, \enquote{Quantum sensing,}
  {\protect\JournalTitle{Rev. Mod. Phys.}} \textbf{89}, 035002 (2017).

\bibitem{Herbschleb2021}
E.~D. Herbschleb, H.~Kato, T.~Makino, \emph{et~al.}, \enquote{Ultra-high
  dynamic range quantum measurement retaining its sensitivity,}
  {\protect\JournalTitle{Nature Communications}} \textbf{12}, 306 (2021).

\bibitem{PhysRevLett.112.103604}
Y.~Israel, S.~Rosen, and Y.~Silberberg, \enquote{Supersensitive polarization
  microscopy using noon states of light,} {\protect\JournalTitle{Phys. Rev.
  Lett.}} \textbf{112}, 103604 (2014).

\bibitem{sciadv2021}
R.~Camphausen, Álvaro Cuevas, L.~Duempelmann, \emph{et~al.}, \enquote{A
  quantum-enhanced wide-field phase imager,} {\protect\JournalTitle{Science
  Advances}} \textbf{7}, eabj2155 (2021).

\bibitem{PhysRevLett.129.150501}
E.~Meyer-Scott, N.~Prasannan, I.~Dhand, \emph{et~al.}, \enquote{Scalable
  generation of multiphoton entangled states by active feed-forward and
  multiplexing,} {\protect\JournalTitle{Phys. Rev. Lett.}} \textbf{129}, 150501
  (2022).

\end{thebibliography}



\section*{Supplementary material}

# Supplemental Material: Dose-efficient Quantum Phase Estimation in Lossy Optical Interferometry

## 1. LOSS MODEL

For the purpose of theoretical optimization, one can equivalently ignore the reference arm altogether and view the problem as a single-mode lossy channel with a fixed mean photon number $N$ entering the sample arm. In this single-mode formulation, the fundamental bound takes the simple form $F_{\mathrm{QL}} = 4\eta N/(1-\eta)$, which can be saturated for large $N$ by suitable displaced, weakly squeezed states, where most of the energy is allocated to the displacement and a sublinear amount to squeezing. The quantum limit for QFI per dose is therefore $\xi_{\mathrm{QL}} \equiv \frac{4\eta}{1-\eta}$.

In particular we consider the general two-mode pure input state $|\psi\rangle_{in} = \sum_{k=0}^{N} \alpha_k |k, N-k\rangle$, where $|k, N-k\rangle$ abbreviates the Fock state $|k\rangle_u |N-k\rangle_d$ with $k$ and $N-k$ photons in the upper and lower arms of the interferometer. Loss is modeled by fictitious beam-splitter of transmissivity $\eta$ in the upper arm of MZI (shown in main text Fig. 1(a)), and cause a Fock state $|k, N-k\rangle$ to evolve into

$$|k, N-k\rangle \rightarrow \sum_{l=0}^{k} \sqrt{B_l^k} |k-l, N-k\rangle \otimes |l\rangle, \tag{S1}$$

where $|l\rangle$ represents the state of one ancillary mode carrying $l$ photons lost from upper mode, while

$$B_l^k = \binom{k}{l} \eta^k (\eta^{-1} - 1)^l. \tag{S2}$$

Including the phase accumulation $|k-l, N-k\rangle \rightarrow e^{i(k-l)\phi} |k-l, N-k\rangle$ and tracing out the ancillary mode results in the output density matrix

$$\rho(\phi) = \bigoplus_{l=0}^{N} p_l |\zeta_l(\phi)\rangle\langle\zeta_l(\phi)|, \tag{S3}$$

where

$$|\zeta_l(\phi)\rangle = \frac{1}{\sqrt{p_l}} \sum_{k=l}^{N} \alpha_k e^{ik\phi} \sqrt{B_l^k} |k-l, N-k\rangle, \tag{S4}$$

is the conditional pure state corresponding to the event when $l$ photons are lost in upper mode, and $p_l$ is the normalization factor corresponding to the probability of that event. Moreover, we have $\langle\zeta_l|\zeta_{l'}\rangle = \delta_{ll'}$. and therefore, we have QFI for the output state

$$F(\rho) = \sum_{l}^{N} p_l F(|\zeta_l(\phi)\rangle\langle\zeta_l(\phi)|) = 4\left( \sum_{k=0}^{N} k^2 x_k - \sum_{l=0}^{N} \frac{(\sum_{k=l}^{N} x_k k B_l^k)^2}{\sum_{k=l}^{N} x_k B_l^k} \right). \tag{S5}$$

An universal upper bound to $F(\rho)$ in terms of the initial state of the probe and any Kraus representation of the quantum channel is introduced in ref [1]. The upper bound is

$$F(\rho) \le C_Q = \frac{4\eta \langle\hat{n}\rangle_0 (N - \langle\hat{n}\rangle_0)}{(1-\eta)(N - \langle\hat{n}\rangle_0) + \eta}. \tag{S6}$$

Here, $\langle\hat{n}\rangle_0 = n_a \le N$ is the average number of photons in the upper arm of the interferometer. This upper bound satisfies $C_Q < F_{\mathrm{QL}}$, and it is obtained when

$$\begin{aligned} |\alpha_0|^2 &= \frac{N - n_a}{N}, \\ |\alpha_N|^2 &= \frac{n_a}{N}, \\ \alpha_k &= 0, \text{ when } k \neq 0 \text{ or } N. \end{aligned} \tag{S7}$$

Thus, it yields the phase estimation uncertainty $\delta\phi \geq \sqrt{1/(\mu C_Q)}$ when the initial state is $|\psi\rangle = \alpha_0|0,N\rangle + \alpha_N|N,0\rangle$ and the experiment is repeated $\mu$ times. The dose is defined as the photons traverse the lossy sample, $d = \sum_k |\alpha_k(\eta)|^2 k = n_a$, where $\alpha_k(\eta)$ are optimal coefficients in Eq. (S7) to reach $C_Q$. Therefore, the corresponding QFI per dose is

$$\xi_Q = \frac{C_Q}{d} = \frac{4\eta}{1-\eta+\frac{\eta}{N-\langle\hat{n}\rangle_0}} \leq \xi_{\mathrm{QL}}. \tag{S8}$$

## 2. PARALLEL STRATEGY WITH UNBALANCED N00N STATES

For $N$-photon unbalanced N00N states of the form $|\psi\rangle_{\mathrm{unb}} = \alpha|N0\rangle + \sqrt{1-\alpha^2}|0N\rangle$, $N$ photons interacts with the sample once, and therefore the dose metric is $d_{\mathrm{unb}} = \alpha^2 N$. After passing the sample, the state evolves as

$$\Lambda(\eta,\phi)|\psi_{\mathrm{unb}}\rangle \otimes |0\rangle = P_0|\zeta_0(\phi)\rangle \otimes |0\rangle + \sum_{l=1}^{N} P_l|\zeta_l\rangle \otimes |l\rangle, \tag{S9}$$

where

$$|\zeta_0(\phi)\rangle = \frac{1}{\sqrt{P_0}}\left(\sqrt{1-\alpha^2}|0N\rangle + \alpha\sqrt{\eta}^N e^{iN\phi}|N0\rangle\right), \tag{S10}$$

$$|\zeta_l\rangle = \frac{1}{\sqrt{P_l}}\alpha\sqrt{B_l^N}|N-l,0\rangle, \tag{S11}$$

$$B_l^N = \binom{N}{l}\eta^N(\eta^{-1}-1)^l, \tag{S12}$$

$$P_0 = 1-\alpha^2+\eta^N\alpha^2, \tag{S13}$$

$$P_l = \alpha^2 B_l^N. \tag{S14}$$

The QFI is given by

$$F_{\mathrm{unb}} = P_0 F(|\zeta_0(\phi)\rangle\langle\zeta_0(\phi)|) + \sum_{l=1}^{N} P_l F(|\zeta_l\rangle\langle\zeta_l|) = 4\eta^N\alpha^2N^2\left(1-\alpha^2\right)/P_{\mathrm{unb}}, \tag{S15}$$

where $P_{\mathrm{unb}} = P_0$ is the probability of detection of all $N$ photons. For all nonzero values of the loss, i.e., $\eta < 1$, $F_{\mathrm{unb}}$ degrades with increasing number $N$ of entangled photons for $N$ sufficiently large. $F_{\mathrm{unb}}$ is close to $\mathcal{C}_Q$ when $\eta \to 1$ or $N \to 1$. We get the QFI per dose $\xi_{\mathrm{unb}} = F_{\mathrm{unb}}/d_{\mathrm{unb}} = 4\eta^N N(1-\alpha^2)/P_{\mathrm{unb}}$, which depends on $\eta$, $N$ and $\alpha^2$. Firstly, we consider $\alpha^2$ to be variant and treat $\eta$ and $N$ as constants, $\xi_{\mathrm{unb}}$ decreases with $\alpha^2$, and

$$\lim_{\alpha^2\to 0}(\xi_{\mathrm{unb}}/\xi_{\mathrm{QL}}) = \eta^{N-1}N(1-\eta) \equiv f(N,\eta). \tag{S16}$$

In this limit, by treating $N$ as a variant and setting the derivative with respect to $N$ to 0, we find $f(N,\eta)$ equals to

$$f(N^*_{\mathrm{unb}},\eta) = (1-1/\eta)/(e\ln\eta), \tag{S17}$$

when $N^*_{\mathrm{unb}} = -1/\ln(\eta)$ for $\eta \in (0,1)$. We observe that $f(N^*_{\mathrm{unb}},\eta)$ decreases as $\eta$ increases. Under highly transmissivity conditions ($\eta \to 1$), the ratio approaches $1/e$ from Eq. (S17). In practice, the number of photons $N$ must be an integer. For a given transmissivity $\eta$, the optimal discrete value is $N_{\mathrm{unb,opt}} = \arg\max_{N\in\{\lfloor N^*_{\mathrm{unb}}\rfloor,\lceil N^*_{\mathrm{unb}}\rceil\}} \xi_{\mathrm{unb}}(N)$, where $\lfloor\cdot\rfloor$, $\lceil\cdot\rceil$ denote the floor and ceiling functions, respectively.

## 3. SINGLE-PASS (SP) STRATEGY

We consider the sequential strategy when the probe is a pure single-photon state $|\psi_{in}\rangle = \alpha_0|01\rangle + \alpha_1|10\rangle$, where $\alpha_1^2 = \alpha^2$ is the transmission coefficient of $BS_1$, and $\alpha_0^2 = 1-\alpha^2$.

For single-pass strategy, the input state interact with the sample once. We have the output density matrix from Eq. (S3), which is

$$\rho(\phi) = P_0|\zeta_0(\phi)\rangle\langle\zeta_0(\phi)| + P_1|\zeta_1\rangle\langle\zeta_1|, \tag{S18}$$

with

$$|\zeta_0(\phi)\rangle = \frac{1}{\sqrt{P_0}}(\alpha_0\sqrt{B_0^0}|01\rangle + \alpha_1 e^{i\phi}\sqrt{B_0^1}|10\rangle), \tag{S19}$$

$$|\zeta_1\rangle = \frac{1}{\sqrt{P_1}}\alpha_1\sqrt{B_1^1}|00\rangle, \tag{S20}$$

and

$$B_0^0 = 1, \tag{S21}$$

$$B_0^1 = \eta, \tag{S22}$$

$$B_1^1 = 1-\eta, \tag{S23}$$

$$P_0 = \alpha_0^2 B_0^0 + \alpha_1^2 B_0^1, \tag{S24}$$

$$P_1 = \alpha^2 B_1^1. \tag{S25}$$

Using the properities of QFI, we get a formula for QFI with explicit dependence on the input state parameters $\alpha_1$,

$$F_{\mathrm{SP}} = 4P_0\left(\left\langle\frac{\partial\zeta_0}{\partial\phi}\middle|\frac{\partial\zeta_0}{\partial\phi}\right\rangle - \left|\left\langle\frac{\partial\zeta_0}{\partial\phi}\middle|\zeta_0\right\rangle\right|^2\right) = 4\left(\eta\alpha_1^2 - \frac{(\eta\alpha_1^2)^2}{\alpha_0^2+\eta\alpha_1^2}\right). \tag{S26}$$

The single photons passing through the sample once, the dose is $d_{SP} = \alpha_1^2$. And therefore the dose efficiency for single pass strategy is $\xi_{\mathrm{SP}} = 4\eta(1-\alpha^2)/(1-\alpha^2+\alpha^2\eta)$, which achieves its maximum $4\eta$ with respect to $\alpha^2 = 0$.

## 4. MULTI-PASS (MP) STRATEGY

For the multi-pass (MP) strategy, as shown in main text Fig. 1(b), the single photons after passing the $BS_1$ is in the superposition state $|\psi_{in}\rangle = \alpha|10\rangle + \sqrt{1-\alpha^2}|01\rangle$. In the $N$-pass sequential strategy, $N$ channels can be applied one after another. Each of the $N$ channels is similar as we here simulate the only sample, which is represent by $\Lambda(\eta,\phi)$, where $\eta$ is the transmissivity coefficient of the sample and $\phi$ is the relative phase shift. The single photon pass the sample, it evolves as $\rho^{(N)}(\phi) = P_0|\zeta_0^{(N)}(\phi)\rangle\langle\zeta_0^{(N)}(\phi)| + P_1|\zeta_1^{(N)}\rangle\langle\zeta_1^{(N)}|$, where $|\zeta_0^{(N)}(\phi)\rangle = 1/\sqrt{P_0}(\alpha\sqrt{\eta^N}e^{iN\phi}|10\rangle + \sqrt{1-\alpha^2}|01\rangle)$, and $|\zeta_1^{(N)}\rangle = 1/\sqrt{P_1}\alpha\sqrt{1-\eta^N}|00\rangle$. The conditional probabilities are $P_0 = \alpha^2\eta^N + 1 - \alpha^2$, and $P_1 = \alpha^2(1-\eta^N)$. Consequently, we get the QFI

$$F_{\mathrm{MP}} = 4P_0[\langle\partial_\phi\zeta_0^{(N)}(\phi)|\partial_\phi\zeta_0^{(N)}(\phi)\rangle - |\langle\partial_\phi\zeta_0^{(N)}(\phi)|\zeta_0^{(N)}(\phi)\rangle|^2] = 4\eta^N\alpha^2N^2\left(1-\frac{\eta^N\alpha^2}{P_0}\right), \tag{S27}$$

and its maximum is achieved when $\alpha^2 = \frac{1}{1+\sqrt{\eta^N}}$ [2]. For the MP strategy, we get the dose injecting into the sample $N$ times as

$$d_{\mathrm{MP}} = \alpha^2\sum_{k=0}^{N-1}\eta^k = \alpha^2\frac{1-\eta^N}{1-\eta}. \tag{S28}$$

And therefore, the QFI per dose of the MP strategy is

$$\xi_{\mathrm{MP}} = \frac{F_{\mathrm{MP}}}{d_{\mathrm{MP}}} = 4\eta^N N^2\frac{1-\eta}{1-\eta^N}\left(1-\frac{\eta^N\alpha^2}{P_{\mathrm{MP}}}\right) \xrightarrow{\alpha\to 0} 4\eta^N N^2\frac{1-\eta}{1-\eta^N}, \tag{S29}$$

with $P_{\mathrm{MP}} = P_0$. In order to find an analytic solution for the optimal number of passes $N$, we permit $N$ to be continuous. $\lim_{\alpha\to 0}\xi_{\mathrm{MP}}/\xi_{\mathrm{QL}} \le \frac{\eta^N N^2}{1-\eta^N}\frac{(1-\eta)^2}{\eta}$, has a single peak over positive values for $N$ and tends to zero as $N$ tends to zero or infinity. By setting the derivative with respect to $N$ to 0, we find $\lim_{\alpha\to 0}\xi_{MP}/\xi_{QL}$ is optimized when $\ln(\eta^N) = 2\eta^N - 2$. We set $\eta^N = -\frac{W}{2}$, and $z_0 = -\frac{2}{e^2}$, the equation becomes $We^W = z_0$, which is exactly the defining equation of the Lambert W function [3, 4]. The principal branch $W \equiv W_0(z_0) \approx -0.406$ at the point $z = z_0$. Hence, the optimal ratio $(\xi_{\mathrm{MP}}/\xi_{\mathrm{QL}})_{\mathrm{opt}} = -W(2+W)(1-\eta)^2/(\eta\ln^2(\eta)) \approx 0.648(1-\eta)^2/(\eta\ln^2(\eta))$

when $\eta^{N^*_{\mathrm{MP}}} \approx 0.203$. Counterintuitively, we have found that in order to lose the fewest photons overall, it is best for the lossy phase to lose $\sim 80\%$. It shows that for measurements of a single phase shift with high transmissivity $\eta \to 1$, MP and parallel strategies produce the same QFI. While considering QFI per dose, MP strategy is better than parallel strategy.

While the analytical solution treating $N$ as a continuous variable provides a clear theoretical bound and captures the essential physical trends, the number of passes $N$ in experimental implementations must be an integer. For a given transmissivity $\eta$, the optimal discrete value is $N_{\mathrm{opt,MP}} = \arg\max_{N \in \{\lfloor N^*_{\mathrm{MP}} \rfloor, \lceil N^*_{\mathrm{MP}} \rceil\}} \xi_{\mathrm{MP}}(N)$.

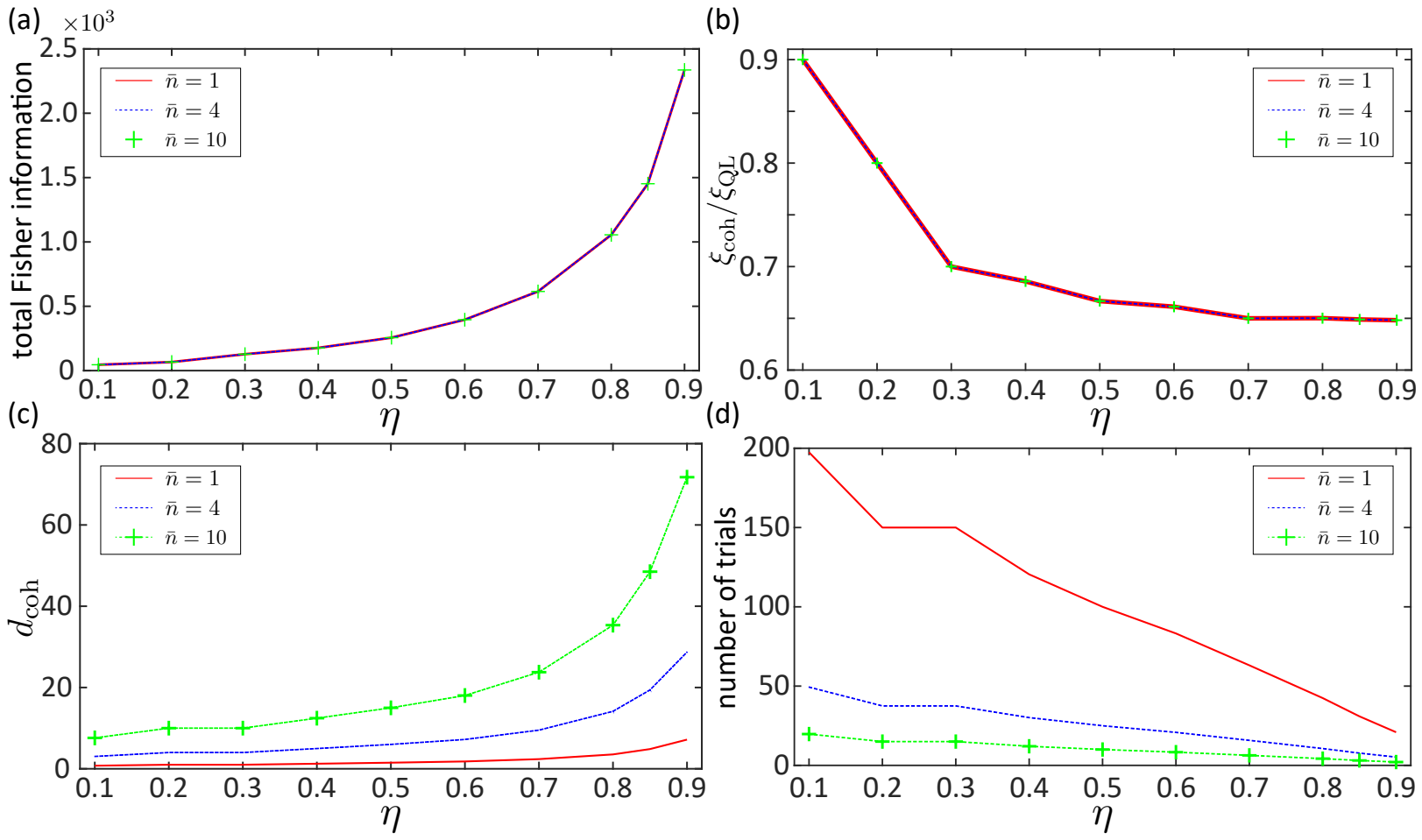


**Fig. S1.** MP strategy with coherent-state probes. Given the total dose constraint $D_{\mathrm{th}} = 150$, (a) the optimized total FI, (b) $\xi_{\mathrm{coh}}/\xi_{\mathrm{QL}}$, (c) the per-trial dose $d_{\mathrm{coh}}$, and (d) the number of experiment trials for different mean photon numbers $\bar{n} = \{1, 4, 10\}$.

The same MP strategy can be applied to coherent-state probes to maximize the Fisher information per dose under realistic loss conditions. We consider a coherent state $|\alpha\rangle$ with average photon number $\bar{n} = |\alpha|^2$ and $N$ passes through the sample. The coherent state $|\alpha\rangle$ is injected into one port of the interferometer and the other input is a vacuum state, the input state is therefore $|0\rangle_a |\alpha\rangle_b$, after passing through the first BS ($\mathrm{BS}_1$ with transmissivity $\sin^2\theta$), the state is $|\psi\rangle = |\alpha \sin\theta\rangle_a |\alpha\cos\theta\rangle_b$. After passing the sample $N$ times, the photon dose is

$$d_{\mathrm{coh}} = \bar{n}\sin^2\theta \sum_{k=0}^{N-1} \eta^k = \frac{\bar{n}\sin^2\theta(1-\eta^N)}{1-\eta}, \tag{S30}$$

where $\sin^2\theta$ describes the transmission of $BS_1$ of the interferometer. Here we use $\sin^2\theta$ rather than the coefficient $\alpha_k$ to avoid confusion, with $\sin\theta = \alpha_1$ and $\cos\theta = \alpha_0$ in the Fock-state case. Using photon counting measurement, and the Fisher information is

$$F_{\mathrm{coh}} = 4\bar{n}\eta^N N^2 \frac{\sin^2\theta\cos^2\theta}{\eta^N\sin^2\theta + \cos^2\theta}. \tag{S31}$$

When $\bar{n} = 1$, it reduces to the single-photon case $F_{\mathrm{MP}}$ and $d_{\mathrm{MP}}$. We obtain the Fisher information normalized by the photon dose,

$$\xi_{\mathrm{coh}} = \frac{F_{\mathrm{coh}}}{d_{\mathrm{coh}}} = \frac{4\eta^N N^2(1-\eta)}{1-\eta^N} \frac{\cos^2\theta}{\eta^N\sin^2\theta + \cos^2\theta}, \tag{S32}$$

which obtains the same FI per dose as in the single-photon case, independent of $\bar{n}$. As the photon number increases, the total dose naturally rises, and photodamage may become relevant. Fig. S1 shows the result of optimizing the coherent-state MP strategy under a dose constraint $D_{\mathrm{th}} = 150$ for different $\bar{n} = \{1, 4, 10\}$. Panels (a) and (b) show that increasing the mean photon number $\bar{n}$ does not change the total Fisher information and the FI per dose, as expected. However, panel (c) shows that the per-trial dose $d_{\mathrm{coh}}$ increases as $\bar{n}$ grows, and panel (d) shows the corresponding number of experiment trials decreases to keep the total dose $D = \mu d = D_{\mathrm{th}}$.

## 5. QFI PER DOSE OF CONTROL-ENHANCED SEQUENTIAL (CS) STRATEGY

In order to describe loss and formally keep the number of particles constant, one needs to introduce an ancillary environmental mode, which is initialized in the vacuum state, $|0\rangle_E$, and $|l\rangle_E$ represents $l$-photon loss. The overall initial density matrix is:

$$\rho_{SE}^{(0)} = \rho_S^{(0)} \otimes |0\rangle_E\langle 0|, \tag{S33}$$

where $\rho_S^{(0)} = |\psi^{(0)}\rangle\langle\psi^{(0)}|$ is the pure input state after the first beam-splitter $BS_1$. $|\psi^{(0)}\rangle = \alpha_0|01\rangle + \alpha_1|10\rangle$, where $\alpha_0 = \sqrt{1-\alpha^2}$, and $\alpha_1 = \alpha$. The lossy sample is modeled as a unitary coupling $\hat{U}_\Lambda$ between the system and the environment. In the single-photon subspace $\{|10\rangle_S, |01\rangle_S\}$, the action of $\hat{U}_\Lambda$ is defined as:

$$\hat{U}_\Lambda(|10\rangle_S|0\rangle_E) = \sqrt{\eta}e^{i\phi}|10\rangle_S|0\rangle_E + \sqrt{1-\eta}|00\rangle_S|1\rangle_E, \tag{S34}$$

$$\hat{U}_\Lambda(|01\rangle_S|0\rangle_E) = |01\rangle_S|0\rangle_E, \tag{S35}$$

where $|00\rangle_S$ denotes the vacuum state of the system, representing the loss of the photon from the interferometric modes. $|00\rangle_S$ is a stationary state under the subsequent evolution; specifically, $\hat{U}_\Lambda(|00\rangle_S|1\rangle_E) = |00\rangle_S|1\rangle_E$. The intermediate control is a unitary operation acting strictly on the system Hilbert space:

$$\hat{\mathcal{T}}_k = \hat{U}_{T_k} \otimes \hat{I}_E. \tag{S36}$$

The intermediate controls act strictly on the system Hilbert space such that $\hat{U}_{T_k}|00\rangle_S = |00\rangle_S$. After $N$ interactions with the sample and $N-1$ intermediate controls, the total unitary evolution $\hat{U}_{\text{tot}}$ is the sequential product:

$$\hat{U}_{\text{tot}} = \hat{U}_\Lambda \hat{\mathcal{T}}_{N-1}\hat{U}_\Lambda \ldots \hat{\mathcal{T}}_1\hat{U}_\Lambda = \hat{U}_\Lambda \prod_{k=N-1}^{1} [\hat{\mathcal{T}}_k \hat{U}_\Lambda]. \tag{S37}$$

where the product is ordered from right to left to follow the chronological sequence of operations. The global state of the system and environment evolves to $\rho_{SE}^{(N)} = \hat{U}_{\text{tot}}\rho_{SE}^{(0)}\hat{U}_{\text{tot}}^\dagger$. Although a photon could be lost to different environmental temporal modes during each pass, any loss event results in the system collapsing to the vacuum sector $|00\rangle_S$. Therefore, the environment can be effectively described by a two-dimensional basis $\{|0\rangle_E, |1\rangle_E\}$, where $|0\rangle_E$ signifies zero loss across all $N$ passes, and $|1\rangle_E$ represents the collective loss sector.

To obtain the reduced state of the system, we perform a partial trace over the environment:

$$\rho_S^{(N)} = \text{Tr}_E[\rho_{SE}^{(N)}] = \sum_l \langle l|\, \hat{U}_{\text{tot}}\, (\rho_S^{(0)} \otimes |0\rangle_E\langle 0|)\, \hat{U}_{\text{tot}}^\dagger\, |l\rangle = \sum_l \hat{M}_l \rho_S^{(0)} \hat{M}_l^\dagger, \tag{S38}$$

where $\hat{M}_l = {}_E\langle l|\hat{U}_{\text{tot}}|0\rangle_E$ ($l = \{0, 1\}$) are the Kraus operators. For $l = 0$, we define the cumulative Kraus operator $\hat{\mathcal{M}}_N \equiv \hat{\mathcal{M}}_{l=0}$ representing the zero-loss trajectory:

$$\hat{\mathcal{M}}_N = {}_E\langle 0|\hat{U}_{\text{tot}}|0\rangle_E = \Lambda(\eta, \phi) \prod_{k=N-1}^{1} [T_k \Lambda(\eta, \phi)]. \tag{S39}$$

In the $\{|10\rangle_S, |01\rangle_S\}$ basis, the single-pass Kraus operator is $\Lambda(\eta, \phi) = \text{diag}(\sqrt{\eta}e^{i\phi}, 1)$. The control operator $T_k$ with transmissivity $t_k$ is defined by the mapping:

$$|01\rangle_S \rightarrow t_k|10\rangle_S - \sqrt{1-t_k^2}|01\rangle_S, \tag{S40}$$

$$|10\rangle_S \rightarrow \sqrt{1-t_k^2}|10\rangle_S + t_k|01\rangle_S. \tag{S41}$$

Considering identical controls, $T_k = T$, the evolution is $\hat{\mathcal{M}}_N = \Lambda(\eta, \phi)[T\Lambda(\eta, \phi)]^{N-1}$. Since any loss event ($l \neq 0$) results in the system collapsing to the vacuum state $|\text{vac}\rangle = |00\rangle_S$, the final density matrix is a mixture of the information-carrying survival state and the vacuum sector:

$$\rho_S^{(N)} = \hat{\mathcal{M}}_N \rho_S^{(0)} \hat{\mathcal{M}}_N^\dagger + (1 - P^{(N)})|\text{vac}\rangle\langle\text{vac}|. \tag{S42}$$

Since the vacuum component carries no phase information, we focus on the post-selected conditional pure state $|\psi^{(N)}\rangle$:

$$|\psi^{(N)}\rangle = \frac{\hat{\mathcal{M}}_N|\psi^{(0)}\rangle}{\sqrt{P^{(N)}}} = \frac{\alpha_1^{(N)}|10\rangle + \alpha_0^{(N)}|01\rangle}{\sqrt{P^{(N)}}}. \tag{S43}$$

The statistical weight of this component, defined as the survival probability $P^{(N)}$, is given by:

$$P^{(N)} = \langle\psi^{(0)}|\hat{\mathcal{M}}_N^\dagger\hat{\mathcal{M}}_N|\psi^{(0)}\rangle = |\alpha_0^{(N)}|^2 + |\alpha_1^{(N)}|^2, \tag{S44}$$

where $\alpha_0^{(N)}$ and $\alpha_1^{(N)}$ are the unnormalized probability amplitude after $N$ passes. This probability $P^{(N)}$ represents the "weight" of the information-carrying component in the mixed state $\rho_S^{(N)}$.

To evaluate the performance of our dose-limited sensing strategy, the ultimate figure of merit is the QFI per dose, defined as $\xi = F_{\mathrm{CS}}/d$. Our analysis reveals that $\xi$ attains its maximum value at the symmetry point $\phi = 0$, establishing it as the optimal operating point for phase estimation. To understand the underlying physical mechanism at this optimal point, we decompose the total QFI for the mixed state $\rho_S^{(N)}$ is decomposed into two distinct components: the classical Fisher information $F_{\mathrm{cl}}$, arising from the variation of the survival probability, and the post-selected pure-state contribution $F_{\mathrm{post}}$:

$$F_{\mathrm{CS}} = F_{\mathrm{cl}} + F_{\mathrm{post}}\,. \tag{S45}$$

By modeling the measurement outcome as a binary variable $l \in \{0,1\}$, where $p_0(\phi) = P^{(N)}$ denotes the photon survival probability and $p_1(\phi) = 1 - P^{(N)}$ represents the probability of photon loss, the classical Fisher information is

$$F_{\mathrm{cl}} = \sum_{l=0,1} p_l(\phi)\Big[\frac{\partial}{\partial\phi}\ln p_l(\phi)\Big]^2 = \frac{[\partial_\phi P^{(N)}]^2}{P^{(N)}[1-P^{(N)}]}\,. \tag{S46}$$

The post-selected contribution is given by

$$F_{\mathrm{post}} = P^{(N)}F(|\psi^{(N)}\rangle) = 4P^{(N)}\left[\Big\langle\frac{\partial\psi^{(N)}}{\partial\phi}\Big|\frac{\partial\psi^{(N)}}{\partial\phi}\Big\rangle - \Big|\Big\langle\frac{\partial\psi^{(N)}}{\partial\phi}\Big|\psi^{(N)}\Big\rangle\Big|^2\right], \tag{S47}$$

where $F(|\psi^{(N)}\rangle)$ denotes the intrinsic QFI of the normalized post-selected pure state. In the context of the CS strategy evaluated at the optimal operating point $\phi = 0$, constructive interference of the probability amplitudes occurs within the sensing arm. This maximizes the expected number of sample interactions, consequently driving the survival probability $P^{(N)}$ to a local minimum. At this extremum, the first-order derivative $\partial_\phi P^{(N)}$ identically vanishes, leading to $F_{\mathrm{cl}} = 0$. As a result, the total Fisher information is entirely defined by the post-selected contribution:

$$F_{\mathrm{CS}} = P^{(N)}\,F(|\psi^{(N)}\rangle). \tag{S48}$$

This indicates that at the optimal working point, the phase sensitivity is solely determined by the quantum state evolution (or phase accumulation) rather than the variations in survival probability caused by loss.

The dose for the CS strategy is

$$d_{\mathrm{CS}} = \sum_{k=0}^{N-1}|\alpha_1^{(k)}|^2, \tag{S49}$$

which lead to the result of QFI per dose $\xi_{\mathrm{CS}} = F_{\mathrm{CS}}/d_{\mathrm{CS}}$. At $\phi = 0$, constructive interference within the sensing arm traps the probability amplitudes, maximizing the photon's effective number of interactions with the sample. Although this maximizes the total dose $d_{\mathrm{CS}}$ in the denominator, the coherent phase accumulation grows quadratically with these interactions. This quadratic enhancement of $F_{\mathrm{post}}$ strictly dominates the linear increase in dose, allowing the ratio $\xi_{\mathrm{CS}}$ to reach its global maximum.

At non-optimal phase points ($\phi \neq 0$), the variation in the survival probability $\partial_\phi P^{(N)} \neq 0$ leads to a non-vanishing classical Fisher information $F_{\mathrm{cl}}$. This contribution physically corresponds to absorptive sensing, where phase shifts are mapped into amplitude losses. However,

bounded by the intrinsic loss scaling, the maximum attainable $F_{\text{cl}}$ at the steepest slope of the interference fringe ($N\phi = \pm\pi/2$) remains substantially lower than the maximum $F_{\text{post}}$ achievable at the symmetry point. Therefore, operating the multi-pass interferometer in this regime ($\phi = 0$), relying entirely on the post-selected quantum phase accumulation, is strictly necessary to approach the ultimate theoretical bound $\xi$.

Hence, for an interested sample with transmissivity $\eta$, we have an $N$-pass CS strategy to measure the unknown phase shift $\phi$, and we can numerically optimize $\xi_{\text{CS}}$ through changing the pass number $N$ and transmissivity coefficient $t_k$ for different controls or $t$ for identical controls.

### A. Total QFI under a fixed dose limit

We consider parameter estimation based on $\mu$ repetitions of the experiment, each consuming a per-trial dose $d$, so that the total photon dose is $D = \mu d$. Under a fixed dose constraint $D \leq D_{\text{th}}$, the total QFI is additive and given by

$$F_{\text{tot}} = \mu F(\eta, d) = \frac{D_{\text{th}}}{d} F(\eta, d) = D_{\text{th}} \xi(\eta, d). \tag{S50}$$

where $F(\eta, d)$ is the per-trial QFI and $\xi(\eta, d) = F(\eta, d)/d$ is the QFI per unit dose. Eq. (S50) shows that maximizing the total QFI under a fixed dose limit $D_{\text{th}}$ is strictly equivalent to optimizing the QFI per dose $\xi$ directly.

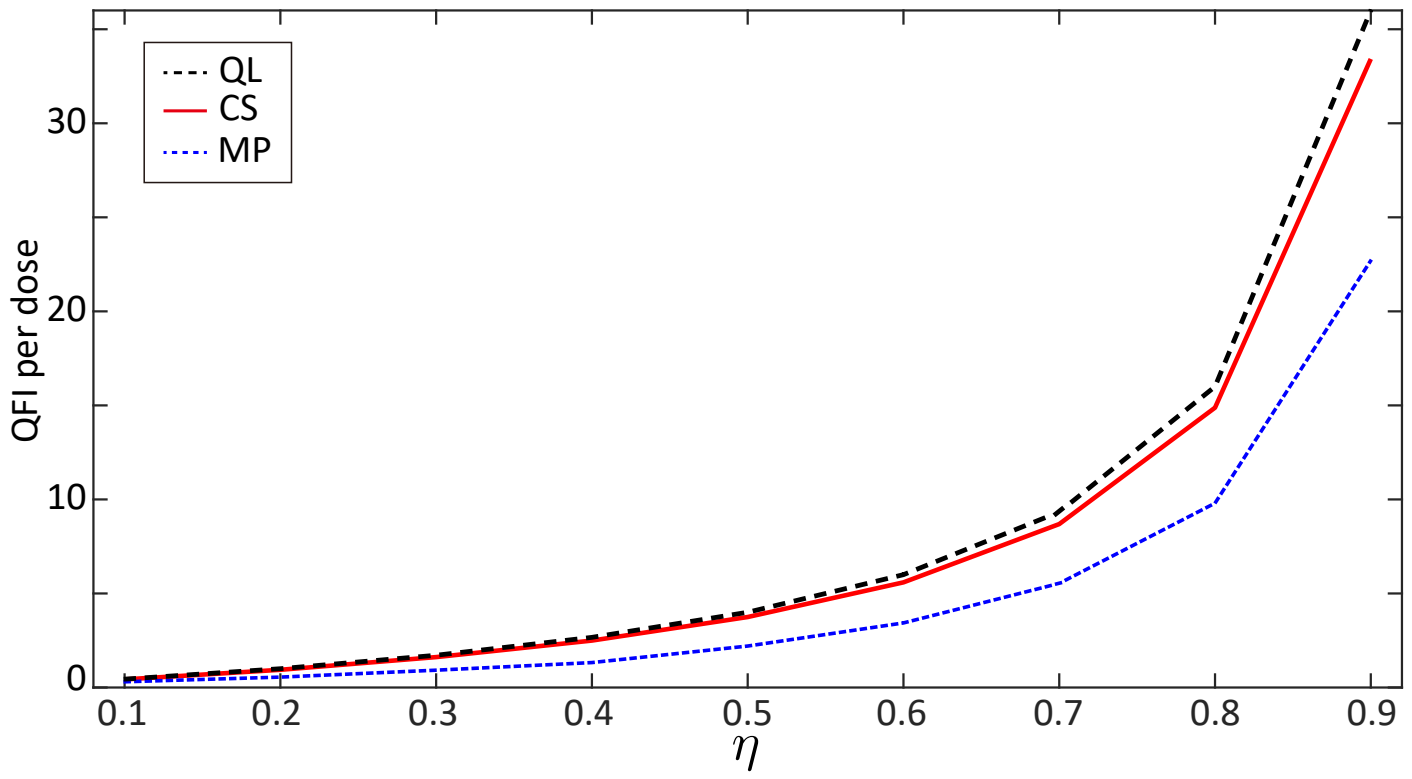


**Fig. S2.** The maximal achievable QFI per dose for both the CS strategy (identical controls) and the MP strategy is plotted against $\eta$.

### B. Optimization of QFI per dose

One can optimize the total QFI in a more direct manner by maximizing the QFI per unit dose,

$$\xi_{\max}(\eta) = \max_{0 \leq d \leq D_{\text{th}}} \frac{F(\eta, d)}{d}. \tag{S51}$$

Since $F_{\text{tot}} = D_{\text{th}}\xi$, this optimization criterion is strictly equivalent to maximizing the total QFI under a fixed dose limit, independent of how the total dose is partitioned between per-trial dose and repetition number. Figure S2 shows the resulting $\xi(\eta)$ for both strategies, together with the quantum-limit benchmark $\xi_{\text{QL}}$. After optimizing the relevant parameters ($\alpha^2, t^2, N$), the QFI per dose is maximized and showing that $\xi_{\text{CS}} > \xi_{\text{MP}}$ across all $\eta$.

Fig. S3 presents a comprehensive comparison between the MP and CS strategies by plotting the ratio of QFI per dose, $\xi_{\text{MP}}/\xi_{\text{CS}}$, as a function of $\eta$ and $N$ for representative initial beam-splitter settings: (a) $\alpha^2 = 0.1$, (b) $\alpha^2 = 0.5$, (c) $\alpha^2 = 0.7$, and (d) $\alpha^2 = 0.9$. We search over pass numbers up to $N \leq 50$ (chosen to balance computational cost and experimental feasibility). For very high sample transmissivity ($\eta \gtrsim 0.9$), the optimum can shift toward larger $N$; nevertheless, the CS formalism and the dose-based optimization procedure remain unchanged. Across the entire parameter region explored, the ratio stays strictly below unity, demonstrating that CS consistently outperforms MP in terms of dose efficiency. Furthermore, Eq. (S29) shows that the QFI per dose for the MP strategy decreases monotonically with increasing $\alpha^2$. We therefore report optimized results at a representative small value, $\alpha^2 = 0.1$, in Tab. S1 (CS strategy with

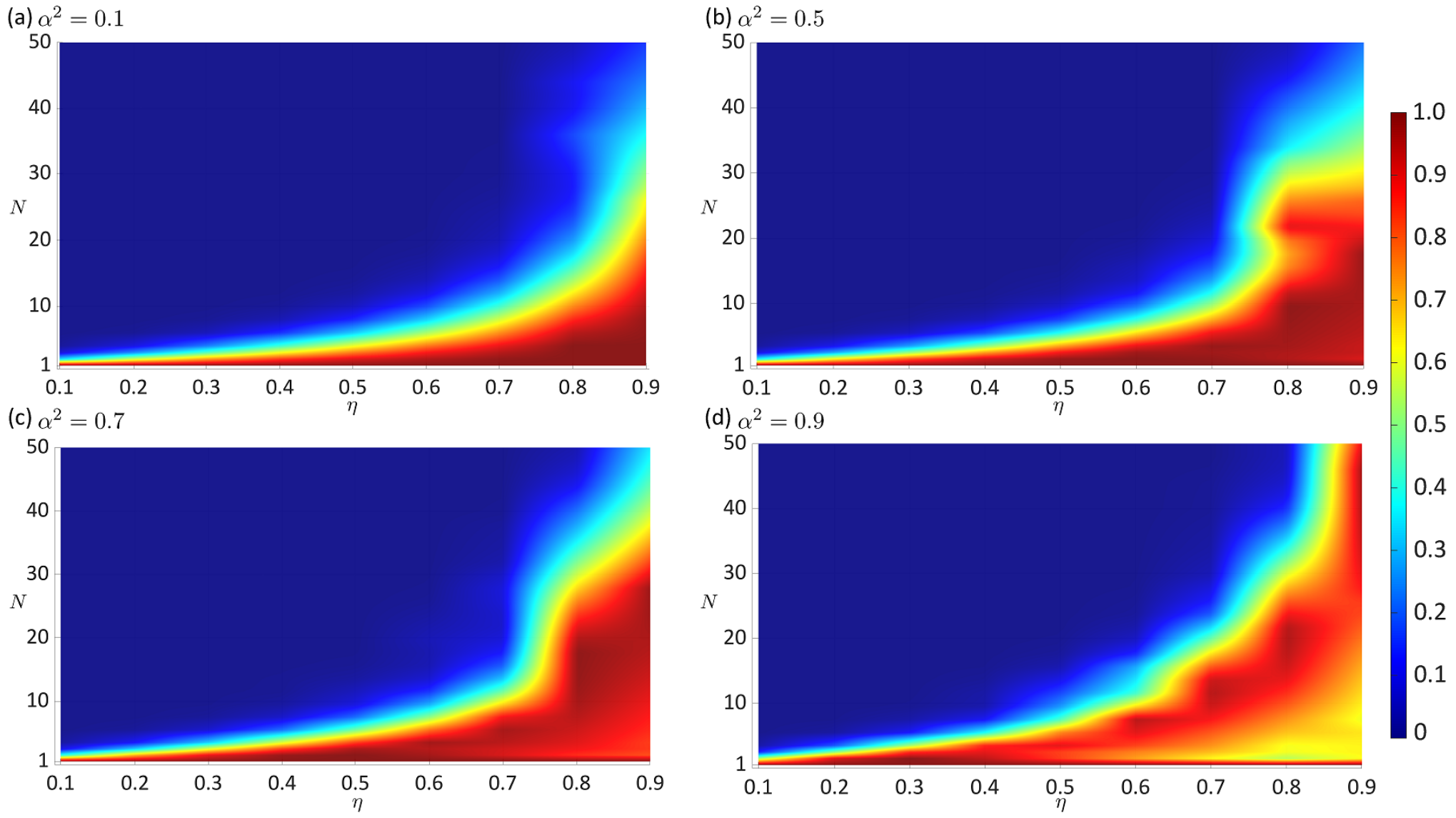


**Fig. S3.** Comparison for QFI per dose of the CS (identical controls) and MP strategies. (a) $\alpha^2 = 0.1$, (b)$\alpha^2 = 0.5$, (c) $\alpha^2 = 0.7$, and (d) $\alpha^2 = 0.9$. The CS strategy surpass MP across all $\eta$ values, demonstrating the advantage of active control in extracting more information per photon through increased effective interactions.

identical controls) and Tab. S2 (CS strategy with different four controls). For all $\eta$ considered, the CS strategy achieves a larger QFI per dose than MP ($\xi_{CS} > \xi_{MP}$), and the corresponding optimal choices of $(\alpha^2, t^2, N)$ differ markedly between the two strategies. Overall, these results show that, within our model and the explored parameter range, active control systematically improves dose efficiency.

Combining the above results, the total QFI under the fixed dose constraint $D_{th}$ is maximized by

$$F_{tot,max} = D_{th}\, \xi_{max}(\eta). \tag{S52}$$

Fig. S4 compares the resulting total QFI for CS and MP strategies with the quantum-limit benchmark $F_{tot,QL} = \xi_{QL} D_{th}$. For all transmissivity $\eta$, the CS strategy achieves a larger total QFI than the MP strategy. Importantly, direct optimization of $\xi$ always yields the maximal achievable total QFI, confirming that QFI per unit dose is the appropriate figure of merit for dose-limited quantum metrology.

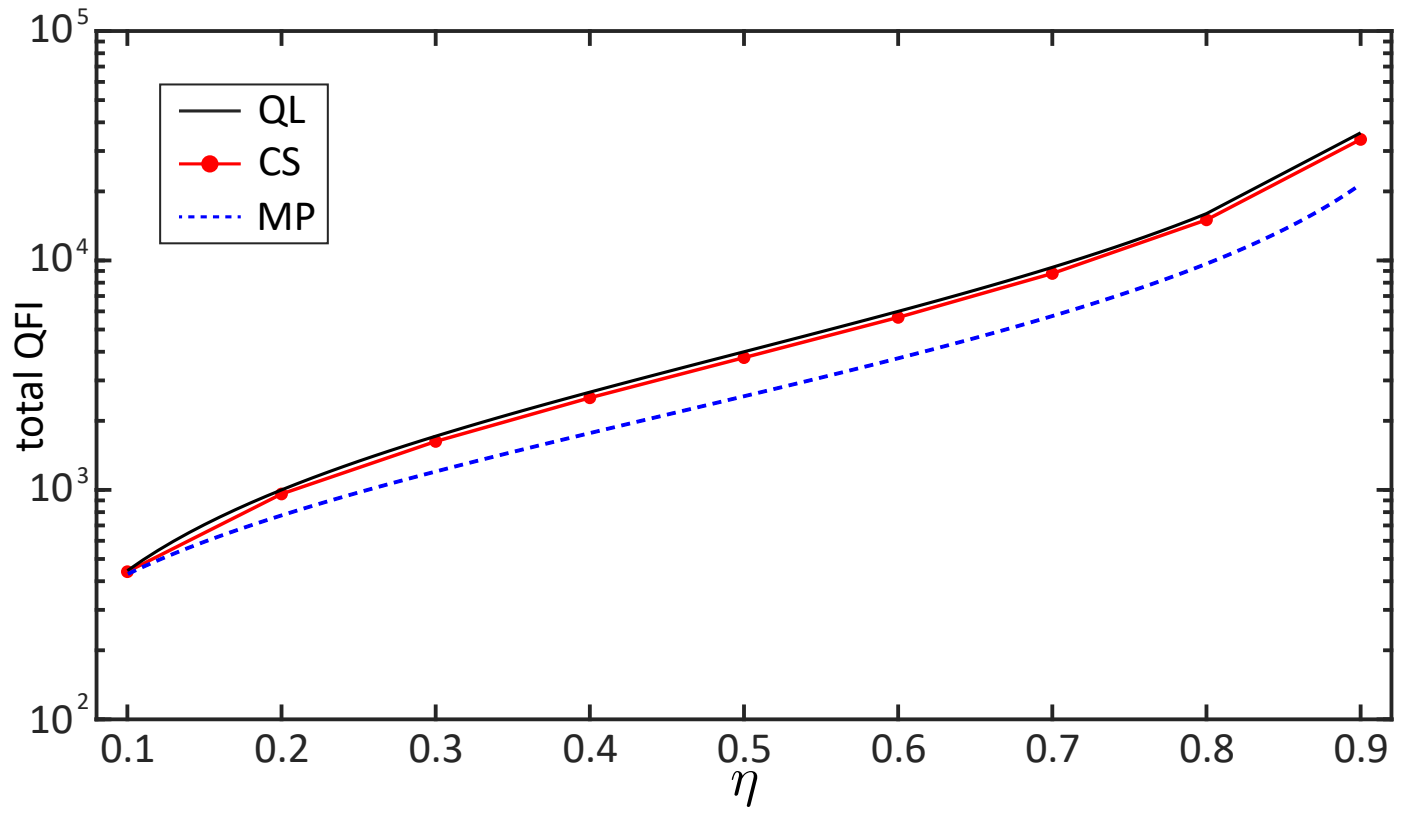


**Fig. S4.** Total QFI for both the CS and MP strategies as a function of $\eta$, under fixed dose constraint $D_{th} = 1000$. The quantum limit of total QFI is $F_{QL} = 4\eta D_{th}/(1-\eta)$.

**Table S1.** **Optimized values of $\xi/\xi_{\mathrm{QL}}$ for a set of $\eta$ in the CS strategy with identical controls and the MP strategy. Set $\alpha^2 = 0.10$.**

| $\eta$ | $\xi_{\mathrm{CS,id}}/\xi_{\mathrm{QL}}$ | $t^2_{\mathrm{opt}}$ | $N_{\mathrm{CS,opt}}$ | $\xi_{\mathrm{MP,N_{CS,opt}}}/\xi_{\mathrm{QL}}$ | $\xi_{\mathrm{MP,N_{MP,opt}}}/\xi_{\mathrm{QL}}$ |
|---|---|---|---|---|---|
| 0.05 | 0.9922 | 0.602 | 2 | 0.1809 | 0.9448,N=1 |
| 0.1 | 0.9796 | 0.393 | 2 | 0.3269 | 0.8901,N=1 |
| 0.15 | 0.9624 | 0.271 | 2 | 0.4424 | 0.8361,N=1 |
| 0.2 | 0.9568 | 0.361 | 4 | 0.0820 | 0.7826,N=1 |
| 0.3 | 0.9465 | 0.224 | 5 | 0.0994 | 0.6774,N=1 |
| 0.4 | 0.9360 | 0.139 | 6 | 0.1332 | 0.6737,N=2 |
| 0.5 | 0.9280 | 0.081 | 8 | 0.1254 | 0.6486,N=2 |
| 0.6 | 0.9186 | 0.042 | 12 | 0.0838 | 0.6457,N=3 |
| 0.7 | 0.9128 | 0.021 | 16 | 0.1097 | 0.6375,N=5 |
| 0.8 | 0.9062 | 0.0083 | 25 | 0.1024 | 0.6353,N=7 |
| 0.9 | 0.8993 | 0.0019 | 50 | 0.1638 | 0.6339,N=16 |
| 0.95 | 0.8420 | 0.0003 | 50 | 0.5820 | 0.6337,N=32 |
| 0.9($\alpha^2 = 0.09$) | 0.9034 | 0.0019 | 50 | / | 0.6353,N=16 |
| 0.95($\alpha^2 = 0.005$) | 0.9035 | 0.0001 | 50 | / | 0.6471,N=31 |

**Table S2.** **Optimized values of $\xi/\xi_{\mathrm{QL}}$ for a set of $\eta$ in the CS strategy with four different controls in the loop. Set $\alpha^2 = 0.10$.**

| $\eta$ | $\xi_{\mathrm{CS}}/\xi_{\mathrm{QL}}$ | $t_1^2$ | $t_2^2$ | $t_3^2$ | $t_4^2$ | $N_{\mathrm{opt}}$ |
|---|---|---|---|---|---|---|
| 0.05 | 0.9946 | 0.60 | 0.80 | / | / | 3 |
| 0.1 | 0.9890 | 0.37 | 0.67 | / | / | 3 |
| 0.15 | 0.9854 | 0.22 | 0.52 | 0.46 | / | 4 |
| 0.2 | 0.9821 | 0.13 | 0.37 | 0.43 | 0.24 | 5 |
| 0.3 | 0.9741 | 0.05 | 0.18 | 0.28 | 0.26 | 5 |
| 0.4 | 0.9560 | 0.08 | 0.10 | 0.16 | 0.22 | 7 |
| 0.5 | 0.9356 | 0.06 | 0.05 | 0.09 | 0.16 | 8 |
| 0.6 | 0.9285 | 0.02 | 0.04 | 0.05 | 0.08 | 11 |
| 0.7 | 0.9195 | 0.01 | 0.01 | 0.02 | 0.06 | 16 |
| 0.8 | 0.9118 | 0.001 | 0.009 | 0.011 | 0.018 | 26 |
| 0.9 | 0.9029 | 0.000 | 0.003 | 0.003 | 0.004 | 50 |

### C. Theoretical upper bound for CS strategy

To derive an analytic upper bound on the QFI per dose, we model the beam-splitter (BS, $T_m = T$ for identical controls) and the sample as $2 \times 2$ matrices:

$$T(t) = \begin{bmatrix} t & r \\ -r & t \end{bmatrix}, \quad S(\phi) = \begin{bmatrix} s(\phi) & 0 \\ 0 & 1 \end{bmatrix}, \tag{S53}$$

where $t$ is the BS amplitude transmissivity, $r = \sqrt{1-t^2}$, and $s(\phi) = \sqrt{\eta}e^{i\phi}$. One sample interaction followed by a BS($T_m$) is then represented by

$$M = T(t)S(\phi) = \begin{bmatrix} ts(\phi) & r \\ -rs(\phi) & t \end{bmatrix}. \tag{S54}$$

This matrix has trace $\mathrm{tr}[M] = t(1+s)$, and determinant $\det[M] = s$. The eigenvalues are therefore

$$\lambda_\pm = \frac{\mathrm{tr}[M] \pm \sqrt{\mathrm{tr}[M]^2 - 4\det[M]}}{2}. \tag{S55}$$

We consider coherent-state probes in the CS strategy. The initial state is $|\psi^{(0)}\rangle = |\alpha t_1\rangle_a |-\alpha r_1\rangle_b$, which can be represented at the level of field amplitudes by the vector $\alpha \begin{bmatrix} t_1 \\ -r_1 \end{bmatrix} = \alpha\, \mathbf{u}_0$, where $t_1$ is the transmissivity of $\mathrm{BS}_1$, with $r_1 = \sqrt{1-t_1^2}$, and $\mathbf{u}_0 \equiv \begin{bmatrix} u_{0,a} \\ u_{0,b} \end{bmatrix}$ denotes the initial amplitude. After N passes through the sample an $N-1$ applications of the $T_m$, the state evolves to

$$|\psi^{(N)}\rangle = S(\phi)M^{N-1}|\psi^{(0)}\rangle = \alpha S(\phi)M^{N-1}\mathbf{u}_0 = \alpha S(\phi)\mathbf{u}_{N-1}. \tag{S56}$$

According to Cayley-Hamilton theorem, any power of the $2 \times 2$ matrix $M$ can be written in closed form as

$$M^n = A_n M + B_n I (n \le 0), \tag{S57}$$

where

$$A_n = \frac{\lambda_+^n - \lambda_-^n}{\lambda_+ - \lambda_-}, \quad B_n = \frac{\lambda_+\lambda_-^n - \lambda_+^n\lambda_-}{\lambda_+ - \lambda_-} = \lambda_+\lambda_- A_{n-1}. \tag{S58}$$

Here $I$ is the identity matrix and $\lambda_\pm$ are the eigenvalues of $M$. Writing the initial field-amplitude vector as $\mathbf{u}_0 = \left[t_1, -r_1\right]^T$, one finds

$$M\mathbf{u}_0 = \begin{bmatrix} t_1 ts - r_1 r \\ -t_1 rs - r_1 t \end{bmatrix}. \tag{S59}$$

Setting $n = N-1$, we obtain

$$M^{N-1}\mathbf{u}_0 = A_{N-1}(M\mathbf{u}_0) + B_{N-1}\mathbf{u}_0. \tag{S60}$$

The final state after $N$ sample interactions can then be written in closed form as

$$|\psi^N\rangle = \begin{bmatrix} \gamma_a(\phi) \\ \gamma_b(\phi) \end{bmatrix} = \alpha \begin{bmatrix} \nu_a \equiv u_{N,a} \\ \nu_b \equiv u_{N,b} \end{bmatrix} = \alpha \begin{bmatrix} s[A_{N-1}(t_1 ts - r_1 r) + B_{N-1}t_1] \\ A_{N-1}(-t_1 rs - r_1 t) - B_{N-1}r_1 \end{bmatrix}. \tag{S61}$$

Because coherent states remain coherent under linear optics and linear loss, and because the system and environment remain in a product of coherent states, tracing out the environment leaves the system in a pure product state,

$$\rho_{out}(\phi) = |\gamma_a(\phi)\rangle\langle\gamma_a(\phi)| \otimes |\gamma_b(\phi)\rangle\langle\gamma_b(\phi)|. \tag{S62}$$

Therefore, the quantum Fisher information is

$$F_Q(\phi) = 4\left(\left|\frac{d\gamma_a}{d\phi}\right|^2 + \left|\frac{d\gamma_b}{d\phi}\right|^2\right) = 4|\alpha|^2(|\nu_a'|^2 + \nu_b'|^2). \tag{S63}$$

The $\phi$-dependence enters only through $s(\phi) = \sqrt{\eta}e^{i\phi}$, with $s'(\phi) = is(\phi)$. For compactness, define

$$X(\phi) \equiv t_1 ts - r_1 r, \quad Y(\phi) \equiv -t_1 rs - r_1 t, \tag{S64}$$

$$X' = t_1 ts' = it_1 ts, \quad Y' = -t_1 rs' = -it_1 rs. \tag{S65}$$

Let $\Delta \equiv \lambda_+ - \lambda_-$ and set $n = N - 1$. Writing $A \equiv A_n$ and $B \equiv B_n$, we have

$$\begin{aligned} A \equiv A_n &= \frac{\lambda_+^n - \lambda_-^n}{\Delta}, \qquad B \equiv B_n = \frac{\lambda_+ \lambda_-^n - \lambda_+^n \lambda_-}{\Delta}. \\ A' &= \frac{n\lambda_+^{n-1}\lambda_+' - n\lambda_-^{n-1}\lambda_-'}{\Delta} + \frac{\lambda_+^n - \lambda_-^n}{\Delta^2}\Delta', \\ B' &= \frac{\lambda_+'\lambda_-^n + \lambda_+ \cdot n\lambda_-^{n-1}\lambda_-' - n\lambda_+^{n-1}\lambda_+'\lambda_- - \lambda_+^n\lambda_-'}{\Delta} + \frac{\lambda_+\lambda_-^n - \lambda_+^n\lambda_-}{\Delta^2}\Delta'. \end{aligned} \tag{S66}$$

It follows that

$$\begin{aligned} \nu_a &= s(AX + Bt_1), \qquad \nu_b = AY - Br_1. \\ \nu_a' &= s'(AX + Bt_1) + s(A'X + AX' + B't_1), \\ \nu_b' &= A'Y + AY' - B'r_1. \end{aligned} \tag{S67}$$

This expression for $F_Q(\phi)$, without invoking any limiting cases, fully captures the combined effects of the loss $\eta$, the beam-splitter parameters $(t_1, t)$, and the pass number $N$ on the achievable phase-estimation precision.

For the CS strategy with a coherent-state probe $|\alpha\rangle$, the dose is

$$d = |\alpha|^2 \sum_{m=0}^{N-1} |u_{m,a}|^2, \tag{S68}$$

where the sensing-arm amplitude at the $m$-th interaction can be written as

$$u_{m,a} = C_+\lambda_+^m + C_-\lambda_-^m. \tag{S69}$$

Using $u_{0,a} = t_1$, and $u_{1,a} = (M\mathbf{u}_0)_a = t_1 ts - r_1 r$, the coefficients are

$$C_+ = \frac{u_{1,a} - \lambda_- u_{0,a}}{\lambda_+ - \lambda_-}, \quad C_+ = \frac{\lambda_+ u_{0,a} - u_{1,a}}{\lambda_+ - \lambda_-}. \tag{S70}$$

It follows that

$$\sum_{m=0}^{N-1} |u_{m,a}|^2 = |C_+|^2 \sum_{m=0}^{N-1} |\lambda_+|^{2m} + |C_-|^2 \sum_{m=0}^{N-1} |\lambda_-|^{2m} + 2\mathrm{Re}\Big(C_+ C_-^* \sum_{m=0}^{N-1} (\lambda_+\lambda_-^*)^m\Big). \tag{S71}$$

with the geometric-series identities

$$\sum_{m=0}^{N-1} |\lambda_\pm|^{2m} = \frac{1 - |\lambda_\pm|^{2N}}{1 - |\lambda_\pm|^2}, \qquad \sum_{m=0}^{N-1} (\lambda_+\lambda_-^*)^m = \frac{1 - (\lambda_+\lambda_-^*)^N}{1 - \lambda_+\lambda_-^*}, \tag{S72}$$

valid for $|\lambda_\pm| \neq 1$ and $\lambda_+\lambda_-^* \neq 1$. Substituting into Eq. (S68) gives

$$d = |\alpha|^2 \left[ |C_+|^2 \frac{1 - |\lambda_+|^{2N}}{1 - |\lambda_+|^2} + |C_-|^2 \frac{1 - |\lambda_-|^{2N}}{1 - |\lambda_-|^2} + 2\mathrm{Re}\Big(C_+ C_-^* \frac{1 - (\lambda_+\lambda_-^*)^N}{1 - \lambda_+\lambda_-^*}\Big) \right]. \tag{S73}$$

We therefore obtain a QFI-per-dose expression that is independent of the mean photon number $\bar{n} = |\alpha|^2$:

$$\xi = \frac{F_Q}{d} = \frac{4(|\nu_a'|^2 + |\nu_b'|^2)}{\sum_{m=0}^{N-1} |u_{m,a}|^2}. \tag{S74}$$

For a fixed loss $\eta$, the maximal value $\xi_{\max}$ can be obtained numerically by optimizing over $(t_1, t, N, \phi)$. To obtain an analytic closed form, we next derive an upper bound on $\xi$, as given below.

At the optimal operating point $\phi = \pi$, the phase-sensitivity amplitudes generated in successive interactions interfere constructively and therefore sum most coherently, yielding the largest

attainable phase response. $s(\pi) = -\sqrt{\eta}$ and $s' = ds/d\phi = is$. We note that this specific operating point is determined by the interplay between the relative phase introduced during state preparation and the algebraic structure of the intermediate controls $T_k$. Because the specific matrix form of $T_k$ routes the probability amplitudes with fixed relative signs, the asymmetric initial state necessitates a phase shift of $\phi = \pi$ to achieve maximal constructive interference in the sensing arm. We confirm that if an initial state with a symmetric relative phase is employed alongside the same $T_k$—for instance, $|\psi^{(0)}\rangle = |\alpha t_1\rangle_a |\alpha r_1\rangle_b$ for coherent probes or $|\psi^{(0)}\rangle = t_1|10\rangle + r_1|01\rangle$ for single-photon probes—the constructive interference condition is modified, shifting the optimal operating point to $\phi = 0$.

Let $|x_k|$ denote the magnitude of the sample-arm amplitude at the $k$-th interaction, with the identification $x_k \equiv u_{m,a}$ $(k-1=m)$. The corresponding dose is

$$d = \sum_{m=0}^{N-1} |u_{m,a}|^2 \equiv \sum_{k=1}^{N} |x_k|^2. \tag{S75}$$

The final state can be represented by the field-amplitude vector

$$\begin{aligned}
\vec{v} &= sM^{N-1}\mathbf{u}_0, \\
\vec{v}\,' &= s'M^{N-1}\mathbf{u}_0 + s\frac{dM^{N-1}}{d\phi}\Big|_{\phi=\pi}\mathbf{u}_0 \\
&= isM^{N-1}\mathbf{u}_0 + s\sum_{k=1}^{N-1} M^{N-1-k}M'M^{k-1}\mathbf{u}_0 \\
&= is\mathbf{u}_{N-1} + s\sum_{k=1}^{N-1} M^{N-1-k}(isP)\mathbf{u}_{k-1} \\
&\equiv i\sum_{k=1}^{N} V_k, \\
V_N &\equiv sM^{N-1}\mathbf{u}_0 = s\mathbf{u}_{N-1} \qquad (\text{for } k=N), \\
V_k &= s^2M^{N-1-k}P\mathbf{u}_{k-1} \qquad (\text{for } k<N), \\
M' &= \begin{bmatrix} ts' & 0 \\ -rs' & 0 \end{bmatrix} = is\begin{bmatrix} t & 0 \\ -r & 0 \end{bmatrix} \equiv isP.
\end{aligned} \tag{S76}$$

Defining the observation row vector $\mathbf{w}^T = [1,\ 1]/\sqrt{2}$, , we can express the QFI at the optimal phase point ($\phi = \pi$) as

$$F_Q = 4\Big(\sum_{k=1}^{N} |\mathcal{A}_k|\Big)^2,\ \mathcal{A}_k = \mathbf{w}^T V_k, \tag{S77}$$

where $V_k$ denotes the (two-mode) phase-response amplitude contributed by the $k$-th interaction. For a vector $\mathbf{y} \in \mathbb{C}^2$,we use the Euclidean norm $||\mathbf{y}|| = (\sum_j |y_j|^2)^{1/2}$, and for a matrix we use the induced (spectral) norm $||A|| = \sup_{||\mathbf{x}||=1} ||A\mathbf{x}||$. Since $M = T(t)S(\phi)$ with $T(t)$ unitary and $||S(\phi)|| = \sqrt{\eta}$, we have

$$||M|| = \sqrt{\eta},\ \ ||M^n|| \le ||M||^n = \eta^{n/2}, \tag{S78}$$

Next, the matrix

$$\begin{aligned}
P\mathbf{u}_{k-1} &= \begin{bmatrix} t & 0 \\ -r & 0 \end{bmatrix}\begin{bmatrix} u_{k-1,a} \\ u_{k-1,b} \end{bmatrix} = \begin{bmatrix} tu_{k-1,a} \\ -ru_{k-1,a} \end{bmatrix}, \\
||P\mathbf{u}_{k-1}|| &\le \sqrt{t^2+r^2}|u_{k-1,a}| = |u_{k-1,a}| \equiv |x_k|, \\
||\mathbf{w}^T|| &= 1,
\end{aligned} \tag{S79}$$

where $|x_k|$ is the magnitude of the sample-arm amplitude at the $k$-th interaction. These norm inequalities yield the following bounds.

$$\begin{aligned}
&\text{for } k=N: A_N = \mathbf{w}^T(s\mathbf{u}_{N-1}), \\
&\qquad |A_N| \le 1\cdot\sqrt{\eta}\cdot||u_{N-1}|| = \sqrt{\eta}|x_N|. \\
&\text{for } k<N: A_k = \mathbf{w}^T(s^2M^{N-1-k}P\mathbf{u}_{k-1}), \\
&\qquad |A_k| \le 1\cdot\eta\cdot||M^{N-1-k}||\cdot||P\mathbf{u}_{k-1}|| \le \eta\eta^{(N-1-k)/2}|x_k| = \sqrt{\eta}\eta^{(N-k)/2}|x_k|.
\end{aligned} \tag{S80}$$

Combining both cases, we obtain the uniform bound

$$|\mathcal{A}_k| \leq \sqrt{\eta}\,\eta^{\frac{N-k}{2}}\,|x_k| \qquad (k = 1, \ldots, N), \tag{S81}$$

and therefore an upper bound on the QFI,

$$\begin{aligned} F_Q &\leq 4\eta \left( \sum_{k=1}^{N} \eta^{\frac{N-k}{2}} |x_k| \right)^2 \\ &\leq 4\eta \left( \sum_{k=1}^{N} \eta^{N-k} \right) \left( \sum_{k=1}^{N} |x_k|^2 \right) = 4\eta \frac{1-\eta^N}{1-\eta} d. \end{aligned} \tag{S82}$$

To ensure the Quantum Fisher Information ($F_Q$) approaches its theoretical limit, the distribution of the sample-arm field amplitudes $|x_k|$ must satisfy very specific conditions. According to the Cauchy-Schwarz inequality $(\sum a_k b_k)^2 \leq (\sum a_k^2)(\sum b_k^2)$, the equality holds if and only if the two vectors involved $a_k, b_k$ are collinear (proportional). This means, the upper bound in Eq. (S82) is reached when

$$b_k \propto a_k \longrightarrow |x_k| = C \cdot \eta^{\frac{N-k}{2}}, \tag{S83}$$

for some constant $C$. Equivalently, the sample-arm intensity (dose per interaction), $n_k = |x_k|^2$, must satisfy $n_k \propto \eta^{N-k}$. Because $\eta < 1$, as $k$ increases (approaching $N$), the term $\eta^{N-k}$ grows larger.

In practice, one does not set each $x_k$ independently; instead, the profile $|x_k|$ is controlled indirectly through the beam-splitter parameters. Specifically, the input splitting $t_1 = x_1$ sets the initial injection into the interferometer, and the inter-pass control parameter $t$ governs how the available dose is redistributed across passes i.e., redistributing $x_k$. In addition to the magnitude requirement on $|x_k|$, reaching the upper bound also requires that all phase-response contributions add coherently at the output, i.e., with the same phase. This constructive addition is achieved by operating at $\phi = \pi$. Under these conditions, the QFI per dose is bounded by

$$\xi \leq \xi_{\text{upp}} = 4\eta \frac{1-\eta^N}{1-\eta}. \tag{S84}$$

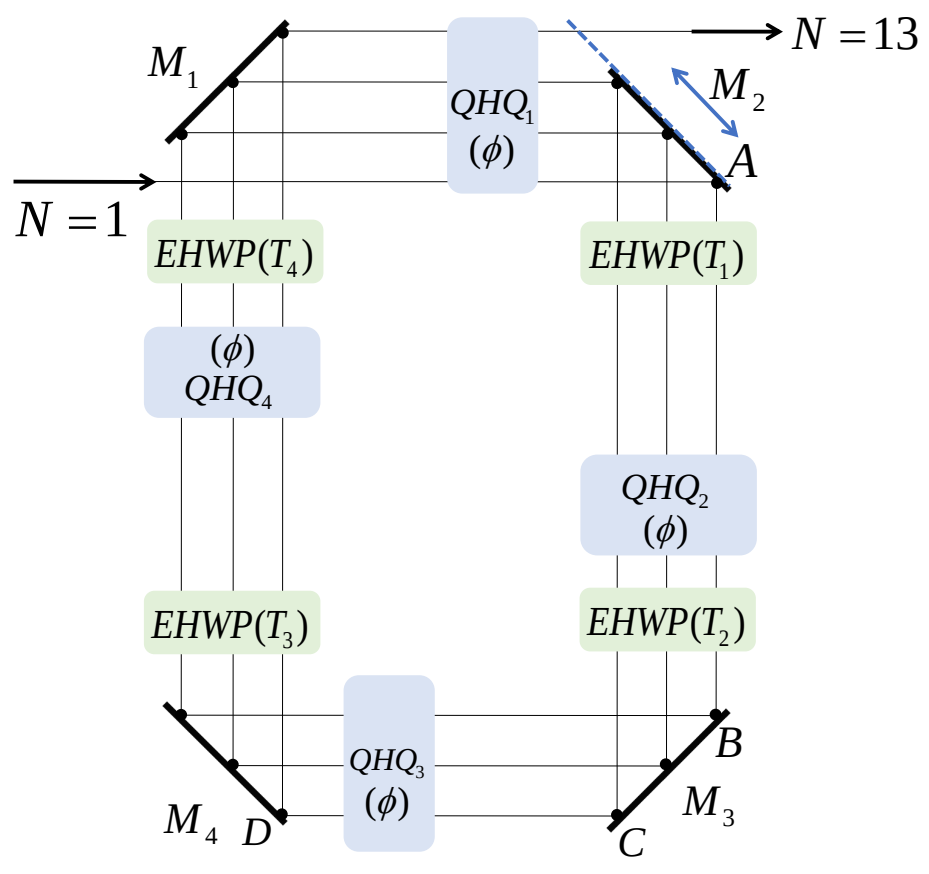


**Fig. S5.** Control the number of pass through a four-mirror loop. The position of $M_{1,3,4}$ is fixed and the position of $M_2$ is adjustable.

## 6. EXPERIMENT DETAILS

Here, we show how to deterministically control the number of passes through the loop in our experiment. We use four mirrors to control the number of passes, denoted as $M_1$ to $M_4$ are laid out as in Fig. S5. The positions of $M_1, M_3$ and $M_4$ are fixed and the number of loops are controlled by slightly moving $M_2$ along the blue dashed line. The maximal pass number $N$ realizable is limited by the size of the clear apertures of optical devices in the loop and the

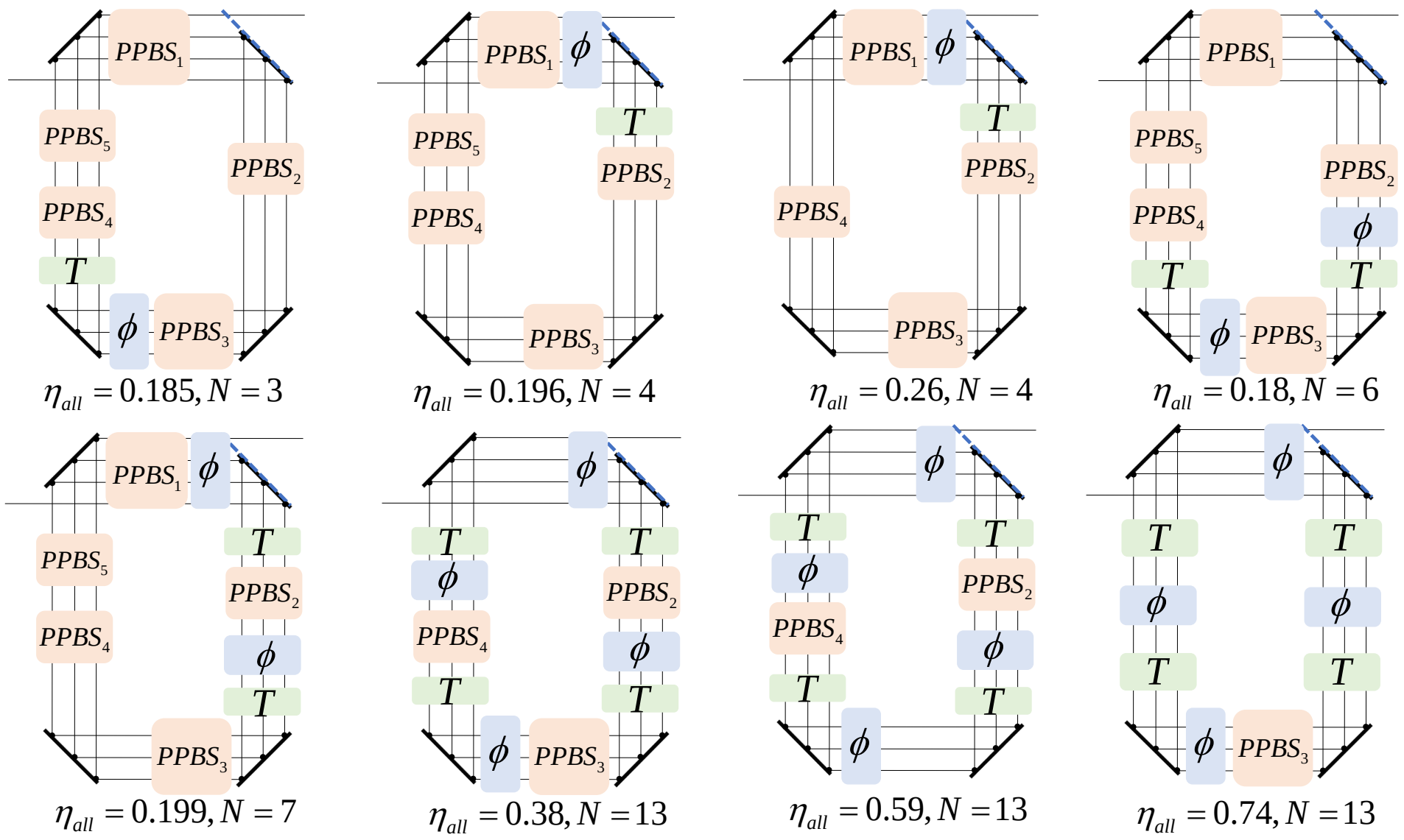


**Fig. S6.** Total transmissivity.

diameter of the beam. In our experiment, the clear apertures of the wave-plates have a diameter of 22 mm. The clear apertures of the mirrors $M_{1-4}$ are 2 inches. The largest radius of the photon beams, which happens for $N = 13$, is around 2.4 mm. The distance between the adjacent beams need to be larger than the diameter of the photon beams while smaller than the diameter of mirrors and the diameter of HWPs. The distance between two adjacent beams in the mirrors are not similar because of the divergence of the photon beams. Thus in the experiment, the diameter of mirrors and HWPs constrains the pass number $N$, avoiding the optical imperfection on the edge. We realize different pass number by gradually moving the position of $M_2$, and therefore the light beam of the final pass is the reflected beam by $M_1$.

We denote the length of A and B in Fig. S5 as $d_{AB}$, which depends on the PPBS-QHQ combination and two EHWPs, and the length of C and D as $d_{CD}$, which depends on the PPBS-QHQ, and both of $d_{AB}$ and $d_{CD}$ are required to be as short as possible to realize possible maximal pass number. We experimentally realize a maximal pass number $N = 13$. Therefore, we put QHQs and EHWPs in the loop with the sequence that $QHQ_3$+$EHWP_3$ ($N = 3$), $QHQ_1$+$EHWP_1$ ($N = 4$), $QHQ_{2,4}$+$EHWP_{2,4}$ ($N = 6$), $QHQ_{1,3}$+$EHWP_{1,3}$ ($N = 7$), and $QHQ_{1,2,3,4}$+$EHWP_{1,2,3,4}$ ($N = 13$).

$\Lambda(\eta)$ represents the single-pass loss introduced by these PPBS components, preparation and measurement module. We use $PPBS_{1-4}$ to simulate the adjustable loss. These PPBSes have the transmissivity $\eta_H = 0.98 - 0.99$ and $\eta_V = 0.93 - 0.95$. We set $N = \{3, 4, 6, 7, 13\}$ with the corresponding sequence of QHQ and EHWP as illustrated before. Additionally, by placing $PPBS_{1/2/3/4}$ in the loop, we get the total relative loss $\eta_{all,N}$. We calibrate $\eta_{all,N} = \eta_{V,N}/\eta_{H,N}$ for each pass $N$ by counting the photons detected in two SPDs for the input state in horizontal and vertical polarization, respectively. And the single-pass transmissivity is $\eta = (\eta_{all,N})^{1/N}$. To calibrate the system's relative loss for horizontal and vertical polarizations, we perform the following steps:

1. Measure System Loss Without PPBS Components:

1) For $N = 13$, put $QHQ_{1-4}$ and $EHWP_{1-4}$ in the loop, setting the HWPs at 0° and the QWPs at 22.5°.

2) Prepare the input state in horizontal polarization, $|H\rangle$.

3) Perform projection measurements on $|H\rangle\langle H|$ and $|V\rangle\langle V|$.

4) Record the average photon counts detected by $D_1$ and $D_2$ over 10 seconds, denoted as $M_{1,H}$ and $M_{2,H}$, respectively.

5) Calculate the horizontal polarization transmissivity:

$$\eta_{no,H} = \frac{M_{2,H}}{M_{1,H} + M_{2,H}}. \tag{S85}$$

6) Repeat the procedure with the input state in vertical polarization, $|V\rangle$, obtaining counts

$M_{1,V}$ and $M_{2,V}$.

7) Calculate the vertical polarization transmissivity:

$$\eta_{no,V} = \frac{M_{2,V}}{M_{1,V} + M_{2,V}}. \tag{S86}$$

8) Determine the relative system loss without PPBS components:

$$\eta_{no} = \frac{\eta_{no,V}}{\eta_{no,H}}. \tag{S87}$$

Through optimization, we achieve $\eta_{no} \approx 0.99$.

2. Measure System Loss with PPBS Components:

1) Insert $PPBS_3$ into the loop.

2) Prepare the input state in horizontal polarization, $|H\rangle$.

3) Record the average photon counts detected by $D_1$ and $D_2$ over 10 seconds, denoted as $M_{1,H}$ and $M_{2,H}$.

4) Calculate the horizontal polarization transmissivity:

$$\eta_{all,H} = \frac{M_{2,H}}{M_{1,H} + M_{2,H}} \approx 0.34. \tag{S88}$$

5) Repeat the procedure with the input state in vertical polarization, $|V\rangle$, obtaining counts $M_{1,V}$ and $M_{2,V}$.

6) Calculate the vertical polarization transmissivity:

$$\eta_{all,V} = \frac{M_{2,V}}{M_{1,V} + M_{2,V}} \approx 0.25. \tag{S89}$$

7) Determine the relative system loss with PPBS components:

$$\eta_{all,N} = \frac{\eta_{all,V}}{\eta_{all,H}} \approx \frac{0.25}{0.34} \approx 0.735. \tag{S90}$$

We calibrate the system's relative loss for each polarization state, ensuring accurate characterization of the PPBS components' impact on the system. We experimentally calibrate the total transmissivity $\eta_{N=3} = 0.18$, $\eta_{N=4} = \{0.196, 0.26\}$, $\eta_{N=6} = 0.18$, $\eta_{N=7} = 0.199$ and $\eta_{N=13} = \{0.38, 0.59, 0.74\}$ as shown in Fig. S6, thus we can get the single-pass transmissivity $\eta = \{0.571, 0.665, 0.707, 0.756, 0.794, 0.928, 0.977\}$.

## 7. COINCIDENCE PROBABILITY

For each value of transmissivity $\eta$, we performed first scan over the entire $2\pi$ range of phase $\phi$ to calibrate the apparatus. The phase dependence of the coincidence counts was used to fit likelihood functions, which includes the imperfections of the state preparation and detection. Next, precise measurements were carried out for at least 20 values of the phases around the sensitive phases with an increment of 0.01 *rad*. For example, we set $N = 13$ passes with $\eta_{N=13} = \{0.38, 0.74\}$ as described in Sec. E. The experiment setup is shown in main text Fig. 2. We prepare the input state $|\psi\rangle = \alpha|10\rangle + \sqrt{1-\alpha^2}|01\rangle$ with $\alpha^2 = 0.95$. The projection measurements are $|H+V\rangle\langle H+V|$ and $|H-V\rangle\langle H-V|$. Firstly, we change the phase $\phi$ with a step 0.1 *rad* over the entire $2\pi$ and observe the superresolution effect to calibrate the QHQ-T combinations in the loop. Secondly, we choose the sensitive phase values $\phi \in (0, \pi/2)$. The interference output is sensitive to the variation of the phase between two arms when the phase is fixed at the optimum value $\pi/(2N) = \pi/26$ and a small phase shift is applied. We change phase values with an increment 0.01 *rad* in the period $\frac{\pi}{13}$. We count the single photons detected in $D_1$ and $D_2$ in 10 seconds, denoted as $M_V$ and $M_H$, respectively. Thus $P_H = M_H/(M_H + M_V)$, and $P_V = 1 - P_H$. The coincidence probabilities at $\eta^{13} = \{0.38, 0.74\}$ as shown in Fig. S7 (a-d) and (e-f), respectively. $P_{H,CS}$ is the red dot, $P_{H,MP}$ is the light green dot, and the fitting results are the yellow and dark green lines. The fitted visibilities are at best $\{0.986, 0.969\}$ and at worst $\{0.964, 0.908\}$ for MP and CS measurement at different phase period.

We use maximum likelihood estimation (MLE) to estimate the phase value. Firstly, we set the experiment for CS/MP strategy and different $\eta$. Then, we scan the phase with a step 0.01 *rad* in period $\pi/N$ for the phase to be measured, as described above. Next, we input $M = 150 - 200$

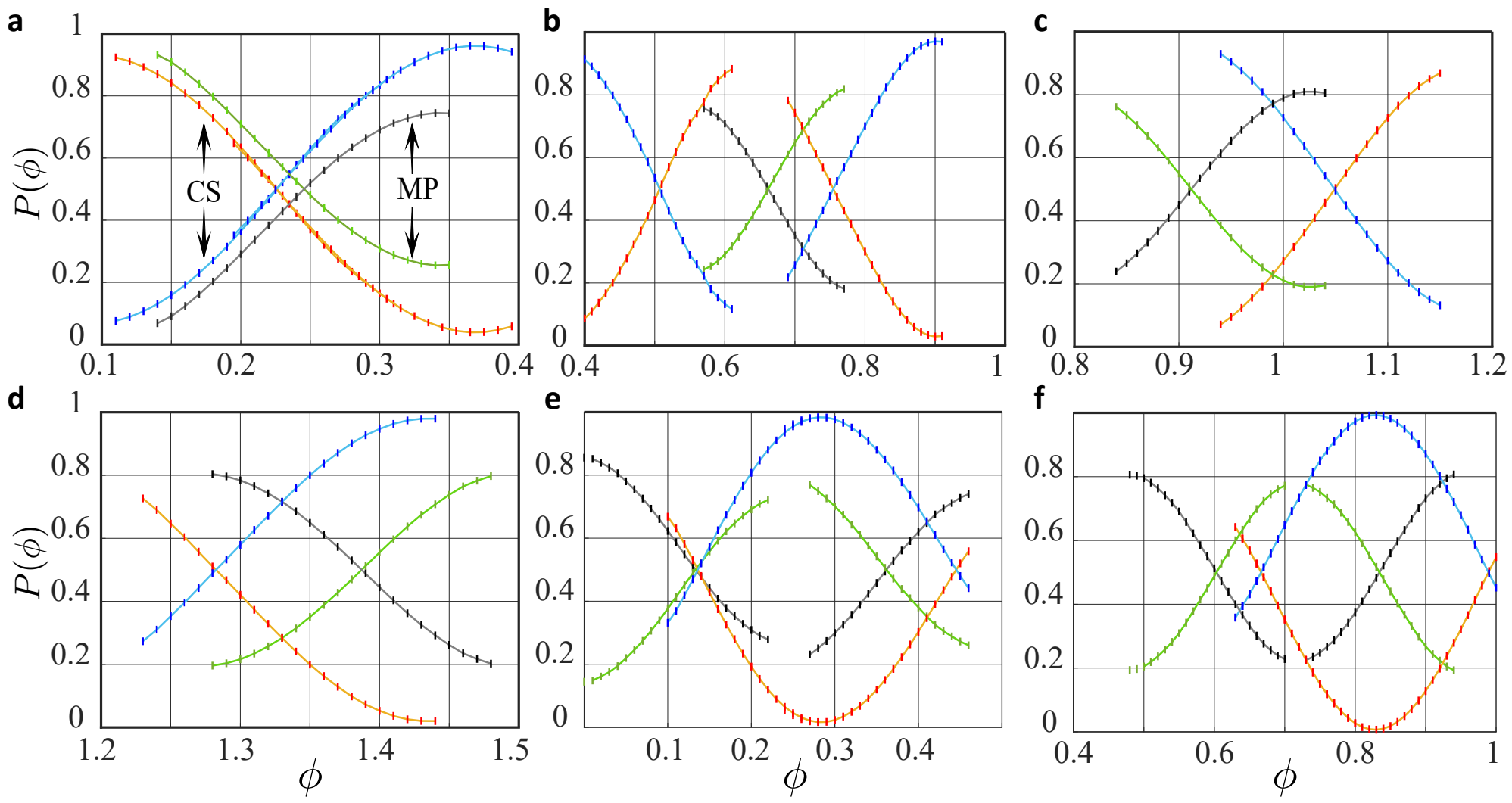


**Fig. S7.** Coincidence probabilities for one period $\frac{2\pi}{13}$ for the strategies: MP strategy experiment data (green and black dots) and the fitting lines, CS strategy experiment data (red and blue dots) and the fitting results lines. In a-d, at total transmissivity $\eta^{13} = 0.380$, the estimated phase values $\phi = \{0.250, 0.665, 0.912, 1.386\}$ for the MP strategy, and $\phi = \{0.223, 0.279, 0.503, 0.758, 0.804, 0.992, 1.056, 1.347\}$ for the CS strategy. In e-f, $\eta^{13} = 0.74$, the estimated phase values $\phi = \{0.131, 0.363, 0.608, 0.829\}$ for the MP strategy and $\phi = \{0.206, 0.342, 0.673, 0.740, 0.899, 1.225, 1.281, 1.462\}$ for the CS strategy.

photons and detect the photons in $\{D_1, D_2\}$, and use MLE to get a estimation value $\hat{\phi}_1$. We repeat the process 50 times to get an mean value $\hat{\phi_2} = mean(\hat{\phi}_1)$ and a variance $va_1 = \delta^2\hat{\phi}_1$. Finally, we repeat the process 10 times and get mean phase value $\hat{\phi} = mean(\hat{\phi}_2)$, and mean value $va = mean(va_1)$ and variance value $\delta^2(va)$ for phase variance of each phase value. Therefore, we can get QFI, $F_j = P_j/(M\delta^2(va))$ $(j = \{\text{CS}, \text{MP}\})$, where $P_j$ is the normalization coefficient of the output state $\psi_j^{(N)}$ of CS/MP strategy. QFI per dose is given by $\xi_j = F_j/d_j$. $d_j$ and $P_j$ are related to $\eta$. For each transmissivity $\eta = \{0.571, 0.665, 0.707, 0.756, 0.794, 0.928, 0.977\}$ and the total dose constraint $D_{\text{th}} = 195$, we set $\alpha^2 = 0.1$ and implement $N = \{3, 4, 6, 7, 13\}$ passes. We estimate the phases and get the results of $\xi/\xi_{\text{QL}}$, as shown in main text Fig.3 a.

For $\eta = \{0.928, 0.977\}$ and dose constraints $D_{\text{th}} = \{1231, 1602\}$, we demonstrate phase estimation uncertainty $\delta\phi = 1/\sqrt{D_{\text{th}}\xi}$ in main text Fig.3 b-c. In our experimental demonstration, the estimated phases deviate from the optimal $\phi = 0$, which introduces a non-zero classical contribution $F_{\text{cl}}$. However, because we estimate the phase using the normalized probabilities of the surviving photons, we inherently discard the photon loss variations. Consequently, our measurement strictly captures the pure-state contribution $F_{\text{post}} = P^{(N)}F(|\psi^{(N)}\rangle)$. Omitting $F_{\text{cl}}$ yields an underestimation of the QFI per dose $\xi_{\text{CS}}$. Notably, even with this underestimation, the CS strategy definitively surpasses both the MP strategy and the parallel benchmark.

## REFERENCES


1. B. Escher, R. L. de Matos Filho, and L. Davidovich, "General framework for estimating the ultimate precision limit in noisy quantum-enhanced metrology," Nat. Phys. **7**, 406–411 (2011).
2. R. Demkowicz-Dobrzanski, U. Dorner, B. Smith, *et al.*, "Quantum phase estimation with lossy interferometers," Phys. Rev. AAtomic, Mol. Opt. Phys. **80**, 013825 (2009).
3. P. M. Birchall, J. L. O'Brien, J. C. Matthews, and H. Cable, "Quantum-classical boundary for precision optical phase estimation," Phys. Rev. A **96**, 062109 (2017).
4. R. M. Corless, G. H. Gonnet, D. E. Hare, *et al.*, "On the lambert w function," Adv. Comput. mathematics **5**, 329–359 (1996).



\end{document}